\documentclass{raa}
\usepackage{graphicx,times}
\usepackage{natbib}
\usepackage{amssymb,amsmath}
\usepackage{hyperref}
\bibpunct{(}{)}{;}{a}{}{,}
\usepackage{url}
\usepackage{amsmath}
\usepackage{mathtools}
\usepackage{upgreek}
\usepackage{footnote}
\usepackage{longtable}

\usepackage{enumitem}
\usepackage{romannum}

\newcommand{\Fg}[1]{Figure \ref{fig:#1}}%beginning of the sentence

\newcommand{\eq}[1]{Eq.~\ref{eq:#1}}
\newcommand{\Eq}[1]{Equation~\ref{eq:#1} }%beginning of the sentence
\newcommand{\eqs}[2]{Eqs.\ \ref{eq:#1} and \ref{eq:#2}}

\newcommand{\se}[1]{Sect.~\ref{sec:#1}}
\newcommand{\Se}[1]{Section~\ref{sec:#1}}%beginning of the sentence
\newcommand{\ses}[2]{Sects.\ \ref{sec:#1} and \ref{sec:#2}}
\newcommand{\Sess}[3]{Section \ \ref{sec:#1}, \ref{sec:#2} and \ref{sec:#3}}

\newcommand{\ie}{i.e.}
\newcommand{\eg}{e.g.}

\newcommand{\AU}{ \ \rm AU}

\newcommand{\Mp}{M_{\rm p}}
\newcommand{\taus}{\tau_{\rm s}}

\newcommand{\alphat}{\alpha_{\rm t}}

\newcommand{\Msyr}{  M_{\odot} \ \rm  yr^{-1}}

\newcommand{\Ms}{  M_{\star}}
\newcommand{\Msun}{  M_{\odot}}
\newcommand{\Me}{ \ M_{\oplus}}

\begin{document}
\pagenumbering{arabic} % page number shown in arabic 
\title{A Tale of Planet Formation: From Dust to Planets}

  \volnopage{{\bf 2020} Vol.~{\bf 20} No.~{\bf 10},~XX(9pp)~
   {\small  doi: 10.1088/1674-4527/20/10/XX}}
   \setcounter{page}{1}         %%starting page, preserved for Editor. DOn't remove!
   \author{Beibei Liu    \inst{1,2}      \and Jianghui Ji \inst{3,4} }

  \institute{ Zhejiang Institute of Modern Physics, Department of Physics \& Zhejiang University-Purple Mountain Observatory Joint Research Center for Astronomy, Zhejiang University, 38 Zheda Road, Hangzhou 310027, China  \\
 \and  Department of Astronomy and Theoretical Physics, Lund Observatory,  Box 43, SE--22100, Sweden
   \\
  \and CAS Key Laboratory of Planetary Sciences, Purple Mountain Observatory, Chinese Academy of Sciences, Nanjing \ 210008, China \\
    \and  CAS Center for Excellence in Comparative Planetology, Hefei \ 230026, China \\ email:  {\it    bbliu@astro.lu.se, jijh@pmo.ac.cn} \\
\vs \no
   {\small Received 2020 August 1; accepted 2020 September 4}
}

\abstract{ 
  The characterization of exoplanets and their birth protoplanetary disks has enormously advanced in the last decade. Benefitting from that, our global understanding of the planet formation processes has been substantially improved. In this review, we first summarize the cutting-edge states of the exoplanet and disk observations. We  further present a comprehensive  panoptic view of modern core accretion planet formation scenarios, including dust growth and radial drift, planetesimal formation by the streaming instability, core growth by planetesimal accretion and pebble accretion. We discuss the key concepts and physical processes in each growth stage and elaborate on the  connections between theoretical studies and observational revelations.  Finally, we point out the critical questions and future directions of planet formation studies. 
\keywords{planets and satellites: general  -- planets and satellites: formation  -- planets and satellites: dynamical evolution and stability  -- protoplanetary disks}
}

   \authorrunning{{\it B. Liu \& J. H. Ji}: From Dust to Planets}            %author_head in even pages
   \titlerunning{{\it B. Liu \& J. H. Ji}:  From Dust to Planets}                 % title_head in odd pages
   \maketitle

%________________________________________________ sections below
%
\section{Introduction}        %% first-level sections will be auto-capitalized
\label{sec:introduction}
In this  article, we review  modern planet formation scenarios in the context of the core accretion paradigm. Since observation and theory are two closely-related aspects, we first recap the detection and characterization of exoplanets in \se{exoplanet} and protoplanetary disks in \se{disk}. The outline of general planet formation processes are given in \se{formation}, classified by the characteristic  sizes of growing  planetary bodies. Finally, we introduce the relevant topics that will be covered in the subsequent sections of the paper.  
\subsection{Exoplanets}
\label{sec:exoplanet}

Half of  the $2019$ Nobel Prize in Physics was awarded to Michel Mayor and Didier Queloz, as an acknowledgement for  their milestone discovery of the first exoplanet orbiting a main-sequence star. This is one of the most influential scientific breakthroughs in astronomy of the past decades. Already in 1995, the above two  astronomers detected the  exoplanet $51$  Pegasi b around  a nearby, Sun-like star in the constellation of Pegasus  \citep{Mayor1995}. Such a discovery was extraordinary and unexpected at that time. It opened an entirely new era in  astronomical observations.  After that, the detection of planets beyond our Solar System has been  enormously developed and grown into a rapidly evolving branch in astronomy. 

One major exoplanet detection method is called radial velocity (RV, or Doppler spectroscopy).  A star and its accompanying  planet co-orbit their center of mass. Observers can see the periodic movement of the star induced by the planet. Due to  the Doppler effect, the observed stellar spectral lines are blueshifted when the star approaches us and are red shifted when the star recedes from us.  Therefore,  the  radial velocity of the star can be acquired  by measuring the displacement  of stellar spectral lines. Through this technique,  the minimum mass of the planet can be obtained.
Since we do not really observe the planet  but infer it  from the wobble of the central star, this is an indirect way to acquire information about the planet. The first exoplanet, $51$  Pegasi b, was discovered by this method. Also, the radial velocity method was involved in most of the  exoplanet discoveries in the early planet-hunting epoch before the launch of the Kepler satellite.

Another leading exoplanet detection method is called transit, which monitors the time variation of a star's brightness to probe the existence of planet(s). When a planet transits in front of its parent star, the surface of the star is partially blocked by the planet and hence the observed stellar flux drops accordingly.  This periodic decrement in the stellar flux reflects the size ratio between the planet and the  star. Therefore, this method can uniquely determine the radius of the planet. Compared to RV that requires  high resolution spectroscopic measurements, transit is a photometric method, and is thus more efficient in detecting planets.  Combining the above two methods together, we can know both the masses and radii, and therefore deduce the bulk densities and chemical compositions of the planets.

 The Kepler satellite is recognized as the most successful planet hunting mission to date, which utilized transit in space to maximize the detection ability and efficiency  \citep{Borucki2010}.
The key  to the success of  the Kepler telescope is that  it has both a large field of view and extremely high photometric precisions. More than $2600$ confirmed exoplanets and $4000$ planet candidates  were detected by Kepler during its nine year operational lifetime ($2009$-$2018$). Thanks to the vastly increased number of exoplanets detected by Kepler, the analysis of planet properties from a statistical perspective has become feasible for the first time \citep{Lissauer2011,Batalha2013,Burke2014}.

\begin{figure}
   \centering
   \includegraphics[scale=0.5, angle=0]{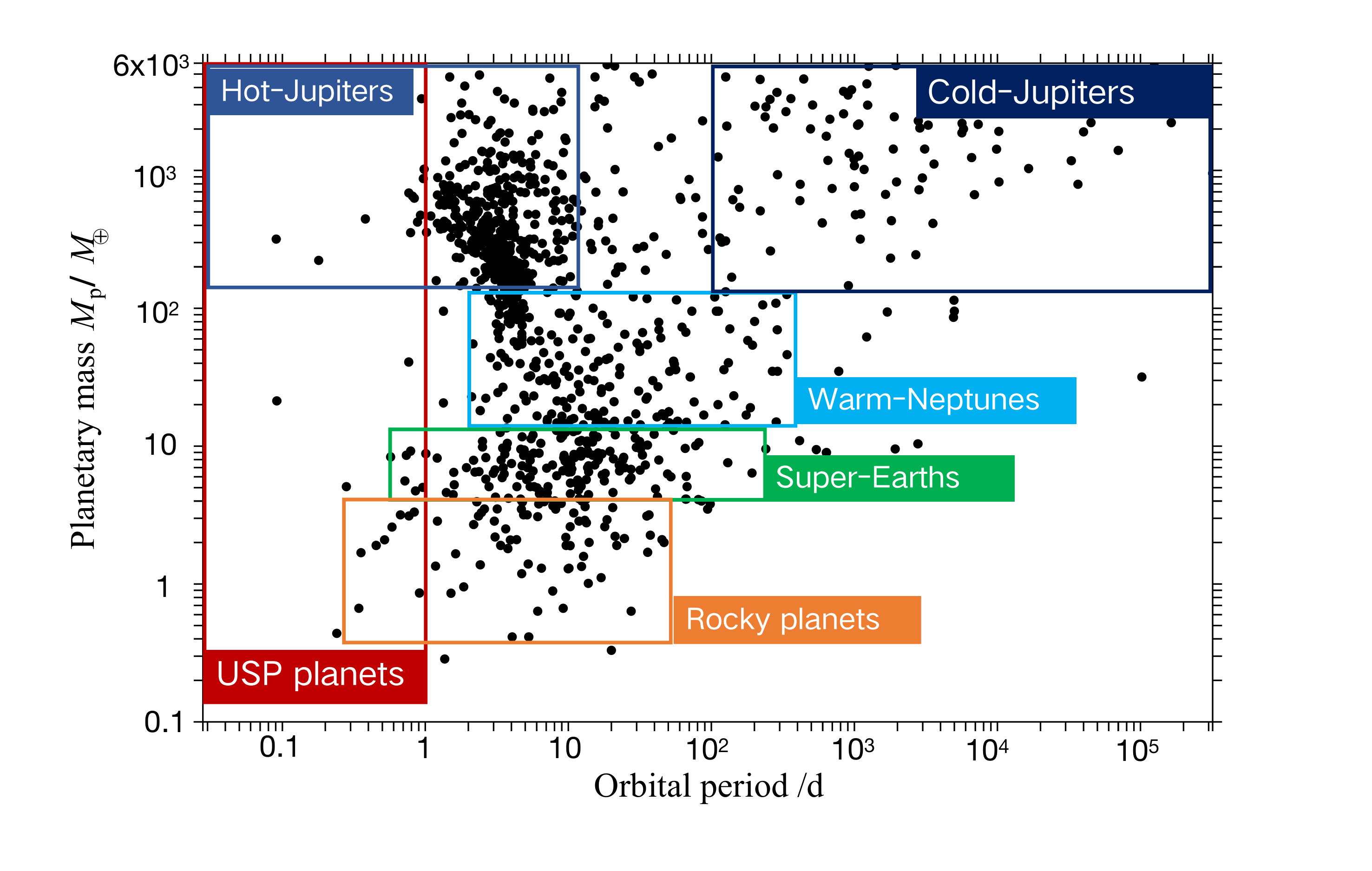}
   \caption{ Orbital period and mass distribution of exoplanet
populations, including hot Jupiters, cold Jupiters, warm-Neptunes, super-Earths and rocky-dominated terrestrial planets. 
A special type of planet with orbital period less than $1$ day is called an ultra-short period (USP) planet. Figure adopted from \cite{Ji2020}. 
   }
   \label{fig:exoplanet}
   \end{figure}

The observed planets are incredibly diverse in terms of masses, sizes, compositions  and orbital properties.  As illustrated in \Fg{exoplanet}, the confirmed exoplanets span several orders of magnitude in their masses and orbital periods \footnote{Data are adopted from \url{http://exoplanet.eu/}}.
  Based on the above two properties, exoplanets can be classified into the following types: 
 hot Jupiters \citep{Mayor1997}, cold Jupiters \citep{ZhuW2018b}, warm (hot) Neptunes \citep{DongS2018}, super-Earths  \citep{Borucki2011} and low-mass rocky planets. 
 \Fg{exoplanet} also marks one particular type of planets with orbital periods less than $1$ day, which are called ultra-short period (USP) planets \citep{SanchisOjeda2014,Winn2018}. Super-Earths strictly refer to the planets with $1.25 {<} R_{\rm p} {<}2 \ R_{\oplus}$  in \Fg{exoplanet}\citep{Borucki2011}. 
 A more general definition of super-Earths is also frequently used in literature studies, referring to planets with radii between  Earth and Neptune  ($1{< }R_{\rm p} {<} 4 \ R_{\oplus}$ or $1{<} \Mp {\lesssim}10 \ M_{\oplus}$,  \citealt{Seager2007,Valencia2007}). With this definition, super-Earths cover the ranges of terrestrial-like, rocky-dominated  planets and sub-Neptune  planets with non-negligible hydrogen envelopes.

 We apply the latter definition of super-Earths in the following discussion of this review. Together with hot and cold Jupiters, these three major types of planets are  currently the  most well characterized and extensively studied samples in literature.  
We summarize  the key observational findings   and the underlying physical interpretations of these three  types of exoplanets (also see reviews of \citealt{ZhouJ2012} and \citealt{Winn2015}).   
\begin{itemize}
	\item Small planets are more common than large planets. When taking into account of the observational bias and for solar-type stars, the occurrence rates of hot Jupiters  and cold Jupiters are  $1\%$ and $5{-}10\%$, respectively, whereas the occurrence rate of super-Earths is $30\%$ \citep{Cumming2008, Howard2010,Mayor2011,Wright2012,DongS2013,Petigura2013,ZhuW2018,Fernandes2019}.  Furthermore, planets are so ubiquitous that they greatly outnumber their host stars \citep{Mulders2018,ZhuW2018}. 
	\item The occurrence rate of giant planets exhibits strong dependences on both stellar mass \citep{Johnson2007,Johnson2010,Jones2016} and metallicity \citep{Santos2004,Fischer2005,Sousa2011}. The occurrence rate of super-Earth seems to be much  more weakly dependent on  stellar metallicity \citep{Sousa2008,Buchhave2012,Buchhave2014,WangJ2015,Schlaufman2015,ZhuW2016,ZhuW2019}. Nevertheless, super-Earths are even more abundant around M-dwarfs compared to those around Sun-like stars \citep{Howard2012,Bonfils2013,Dressing2015,Mulders2015,YangJY2020}. 
		\item Hot Jupiters are nearly circular while the mean eccentricity of cold Jupiters is ${\sim}0.25$ \citep{Marcy2005}. Obliquity defines the angle between the spin axis of the host star and the orbital angular momentum axis of the planet, which can be measured through the Rossiter-McLaughlin effect \citep{Winn2010}. Many hot Jupiters manifest high obliquities, sometimes even polar or retrograde \citep{Triaud2010,Albrecht2012}. Theoretically, hot Jupiters were proposed to grow at further out disk locations and then migrate inward to the present-day orbits \citep{Lin1996}.  Planet-disk interaction will lead the giant planets on circular and coplanar orbits \citep{Lin1993,Artymowicz1993,Ward1997,Nelson2000}, while  the high eccentricities and inclinations of giant planets can originate from a Kozai-Lidov cycle \citep{Kozai1962,Lidov1962} induced by a distant companion  \citep{ Wu2003,Fabrycky2007,Naoz2011,Naoz2013,DongS2014,Anderson2016} or planet-planet scatterings \citep{Rasio1996,Chatterjee2008,Juric2008,Ford2008,Dawson2013}. The latter two planetary  dynamical processes could result in the observed high obliquities. On the other hand, such misalignments could also arise from re-orientation of the host star's spin through internal waves  \citep{Rogers2012,Lai2012} or tilted protoplanetary disks through  binary-disk interactions \citep{Lai2014,Matsakos2017,Zanazzi2018}.		
\item Based on  Kepler data, multi-transit systems have relatively low eccentricities and inclinations while single-transit systems exhibit much higher eccentricities and inclinations \citep{Tremaine2012,Johansen2012,Fabrycky2014,Xie2016,ZhuW2018}.
One hypothesis is that these single-transit planets  come from multiple systems. Their orbits are further excited/disrupted by long-term planet-planet interactions or by outer companions, causing them to appear as ``singles" in transit surveys \citep{Pu2015,Lai2017,Mustill2017}. 
	\item  Hot Jupiters/Neptunes seldom have nearby companions up to a few AUs \citep{Steffen2012,DongS2018}, consistent with  the Kozai-Lidov cycle and planet-planet scattering scenarios \citep{Mustill2015}. On the contrary, nearly half of  warm Jupiters co-exist with low-mass planets \citep{HuangX2016}. Cold Jupiters also seem to be commonly accompanied  by close-in, super-Earths \citep{ZhuW2018b,Bryan2019}. Besides, $25\%{-}30\%$ of the systems with a cold Jupiter are found to host additional giant planets  \citep{Wright2009,Wittenmyer2020}. 
	\item The period ratios of adjacent planet pairs neither show strong pile-ups at mean motion resonances (MMRs) nor uniform distribution in Kepler data. These  planets exhibit an asymmetric distribution around major resonances, such as $2$:$1$ and $3$:$2$ MMRs (Figure 6 of \citealt{Winn2015}). Different scenarios are proposed to explain the above features, including tidal damping \citep{Lithwick2012,Batygin2013,Lee2013,Delisle2014,Xie2014}, retreat of the inner magnetospheric cavity \citep{Liu2017a,Liu2017b}, resonant overstability \citep{Goldreich2014}, interaction with planetesimals \citep{Chatterjee2015}, stochastic migration in highly turbulent disks \citep{Rein2012,Batygin2017} or in shock-generated inviscid disks \citep{McNally2019,YuC2010},  mass growth of a planet \citep{Petrovich2013,WangS2017},  and dynamical instability of tightly packed planetary chains \citep{Izidoro2017,Izidoro2019, Ogihara2015,Ogihara2018}.
		\item The occurrence rate of close-in super-Earths has a bimodal radius distribution, with a factor of two drop  at  $R_{\rm p} {\sim} 1.5{-}2 R_{\oplus}$ \citep{Fulton2017,Fulton2018,VanEylen2018}. 
		This so-called planetary radius valley implies a composition transition from rocky planets without $\rm H$/$\rm He$ gaseous envelopes to  planets with envelopes of a few percent in mass \citep{Lopez2013,Owen2013}.  The above radius gap can be explained by the gas mass loss due to  stellar photoevaporation \citep{Owen2017,Jin2018} or core-powered heating \citep{Ginzburg2018,Gupta2019}. Besides, giant impacts may also contribute to this compositional diversity by striping the planetary primordial atmospheres through  disruptive collisions \citep{LiuSF2015,Inamdar2016}.
Based on the photoevaporation model,   \citep{Owen2017} deduced  that the composition of these super-Earths are rocky dominated, ruling out  low-density, water-world planets. Although this interpretation should be taken with caution, the water-deficit outcome may be caused by the fact that  short-lived radionuclides dehydrate planetesimals during their early accretion phase \citep{Lichtenberg2019},  thermal effects  take place in planetary interiors during long-term evolution phase \citep{Vazan2018}, or the planets experience a runaway greenhouse effect and lose substantial surface water through photo-dissociation \citep{Luger2015,Tian2015}. 
\end{itemize} 

The above results  are summarized based on the current demographic and orbital properties of exoplanets For  knowledge of exoplanet atmospheres, we recommend a recent review by \cite{ZhangX2020}. \Fg{mission} exhibits the launched and future planned space missions for exoplanet detection and characterization from the National Aeronautics and Space Administration (NASA), European Space Agency (ESA) and China National Space Administration/Chinese Academy of Sciences (CNSA/CAS). For instance,  the successor of the Kepler mission, Transiting Exoplanet Survey Satellite (TESS) which was launched in $2018$, aims at discovering short period planets around nearby stars  \citep{Ricker2015,HuangX2018}. Compared to Kepler, the advantage of TESS is that the target stars are easier for ground-based and space-based  follow-up characterization observations.

On the other hand, three Chinese space missions have been initiated and  approved for  detection of exoplanets in the coming decades.
The Chinese Space Station Telescope (CSST), scheduled for launch in $2024$,  will survey mature Jupiter-like planets, Neptunes and super-Earths around solar-type stars using a high-contrast imaging technique, which expects to discover tens of exoplanetary candidates and brown dwarfs.  
The Closeby Habitable Exoplanet Survey (CHES) mission aims at searching for terrestrial planets in habitable zones around  solar-type stars within $10$  pc by using astrometry in space. CHES will observe the target stars with high astrometric precisions of microarcsecond at the Sun-Earth L$2$ point. The mission expects to discover at least $50$ Earth-like planets or super-Earths around $100$ FGK stars with well-determined masses and orbital parameters.
Miyin, on the other hand,  is designed for detecting  habitable exoplanets around nearby stars with interferometry. To achieve direct imaging of these exoplanets and assess their habitability, the mission will launch spacecrafts with groups of telescopes working in the mid-infrared wavelengths, which ensures an extremely-high spatial resolution of $0.01$ arcsecond.

 \begin{figure}
   \centering
   \includegraphics[scale=0.35, angle=0]{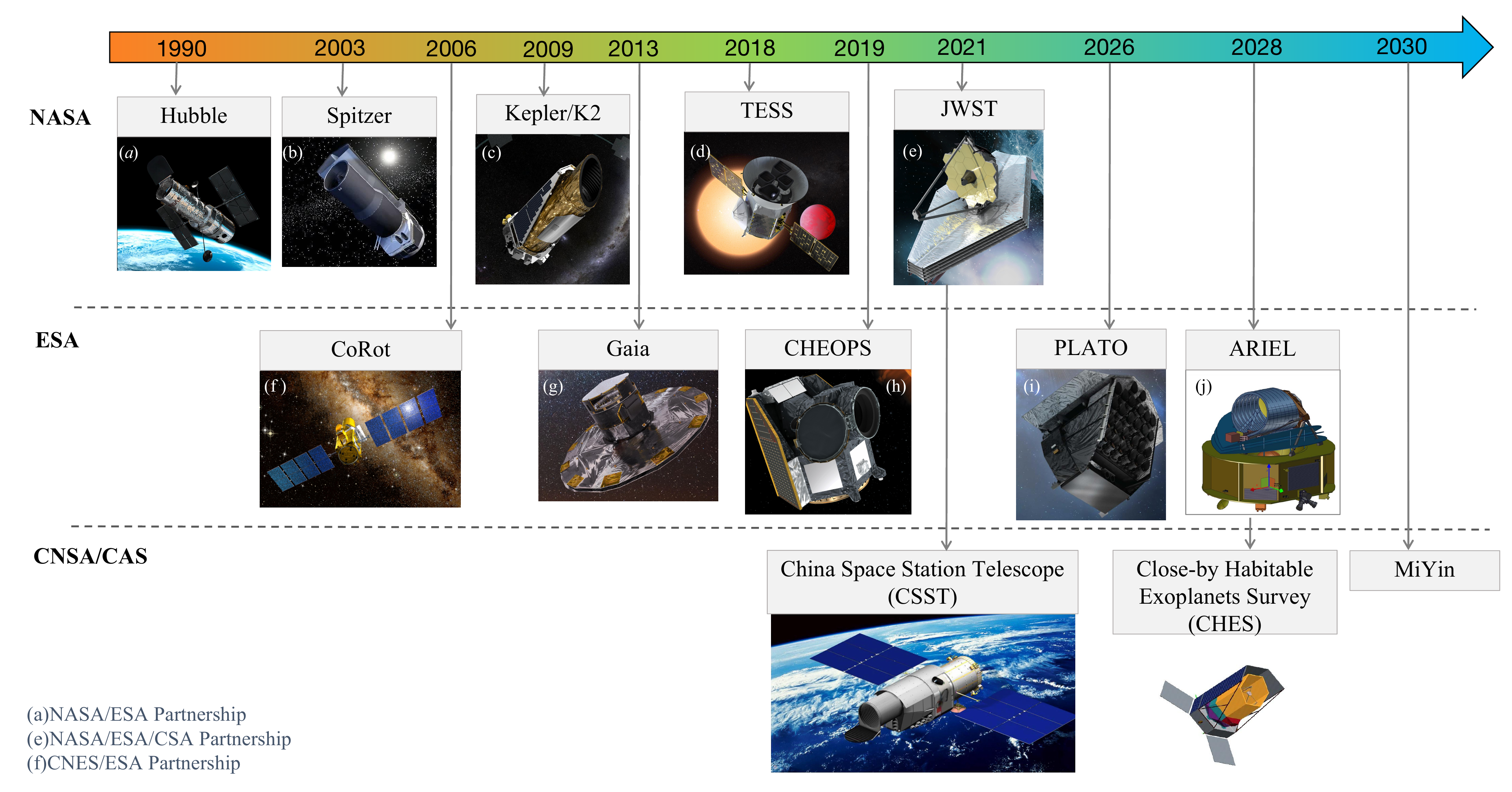}
   \caption{ Launched and future-designed exoplanets exploration missions from NASA, ESA and CNSA/CAS. Credit: (a–e) NASA; (f-j) ESA. Figure adopted from \cite{Ji2020}. 
   }
   \label{fig:mission}
   \end{figure}

\subsection{Protoplanetary disk observation}
\label{sec:disk}

Planets form in protoplanetary disks surrounding their infant stars.  Since the birth and growth of planets are tightly related with their forming environment, studying the physical and chemical conditions of protoplantary disks becomes essential to understanding the planet formation processes.  A large number of young protoplanetary disks have been observed and  extensively studied  in literature.   Here we briefly introduce the up-to-date  observations of disk substructures and provide several hints for the existence of emerging planets. Discussions on the disk solid masses and dust sizes will be presented in \se{pebble_obs}.

Thanks to the unprecedentedly high sensitivity and angular resolution of the Atacama Large Millimeter/submillimeter Array (ALMA), we now have  the capability to reveal the disk structures at  a spectacular level of detail. The spatial resolution of the resolved disks in nearby star forming regions is approximately $3{-}5$ AU.
Axisymmetric rings, gaps, inner cavities and spirals are commonly observed among these disks over a wide range of ages and masses of stellar hosts \citep{Andrews2018,HuangJ2018,Long2018}.  These  disk substructures  are in disagreement with the traditional picture of a smooth disk profile.  For instance, \Fg{disk}(a) shows the disk of HL Tau with a series of concentric bright rings separated by faint gaps from the dust continuum emission \citep{ALMA2015}.

There are different explanations for the formation of the above ring-like substructures. These features can be explained  by grain growth \citep{Zhang2015} or dust  sintering  \citep{Okuzumi2016} at the condensation fronts of major volatile species,  zonal flows in magnetized disks \citep{Flock2015}, a combined effect of the above mechanisms \citep{Hu2019} or secular gravitational instability \citep{Takahashi2014}.  Apart from those interpretations, the most widely accepted scenario is that these substructures are induced by gap-opening planets \citep{Pinilla2012b,Dipierro2015,DongR2015a,Jin2016,
Fedele2017,LiuSF2018,ZhangS2018,LiuY2019,Eriksson2020}. One important note is that,  if these ring-like features are indeed the planet origin, the young age of HL Tau ($t{<}1$ Myr) implies that planet formation may be faster than previously thought. In addition, this early formation hypothesis may also be supported by \cite{Harsono2018}, who analyzed the radial distributions of disk dust and gas around  $\rm TMC1A$ and suggested that millimeter-sized grains have already formed around  such a young Class I object at an age of  ${\sim}10^{5}$ yr. 

In addition to the disk dust-component,  the properties of gas, such as gas velocities,  can also be obtained from molecular line emissions of  
 $\rm CO$ isotopes. The kinematic deviations of gas velocities from Keplerian flows, together with the detected dust gaps at the same disk locations, strongly support the existence of embedded planets associated with gap-opening \citep{Teague2018,Pinte2018,Pinte2020}.

The revealed spiral structures in disks from the scattered light images  have been proposed to feature different origins as well \citep{Muto2012,Grady2012,Stolker2016,Benisty2016,MuroArena2020}. For instance, these patterns can be explained by  density waves excited by the planets \citep{ZhuZ2015b,DongR2015b,Fung2015,Bae2018}. In the early phase when the disk is massive and self-gravitating, the gravitational instability can also induce large-scale spiral arms \citep{Lodato2005,DongR2015c}. In addition, it might also be caused by the shadow from the warped disk \citep{Montesinos2016}, or a Rossby wave instability triggered  vortex \citep{LiH2000,HuangP2019}.

Due to diverse explanations,  previously mentioned  disk kinematics could be treated as  indirect indications of the planets. More straightforward evidence of young planets and  ongoing planet formation is demonstrated as follows. The first clue comes from the RV measurements of young stars, where hot Jupiter candidates are reported around  CI Tau \citep{JohnsKrull2016} and V$830$ Tau \citep{Donati2016}.  \footnote{\cite{Donati2020} pointed out that the RV modulations of CI Tau may also be attributed to the stellar activity.}
 Furthermore, \cite{Plavchan2020} discovered a Neptune-sized planet co-existing with a debris disk around the nearby M dwarf star $\rm AU \  Mic$ by transit surveys. All above stars are in their pre-main-sequences with an age of approximately  $10$ Myr.  
In addition, two embedded planets have been  detected in PDS $70$'s protoplanetary disk by using  the high-contrast imager Spectro-Polarimetric High-contrast Exoplanet REsearch (SPHERE) on European Southern Observatory's  Very Large Telescope (ESO's VLT, \citealt{Keppler2018,Muller2018}). \Fg{disk}(b) displays a synthetic image of the PDS $70$ system, where two planets reside inside the gap of their protoplanetary disk.  This is the first time that young planets have been directly imaged in their birth environment. 
Further analyses with submillimeter continuum  and resolved  H$\alpha$ line  emissions  indicated the presence of a circumplanetary disk \citep{Isella2019} as well as the proceeding gas accretion onto planet  \citep{Haffert2019}.
All these findings, together with the previous results from the disk morphologies/kinematics,  provide valuable constraints on how, when and where planets can form. 
    \begin{figure}
   \centering     
    \includegraphics[scale=0.435,angle=0]{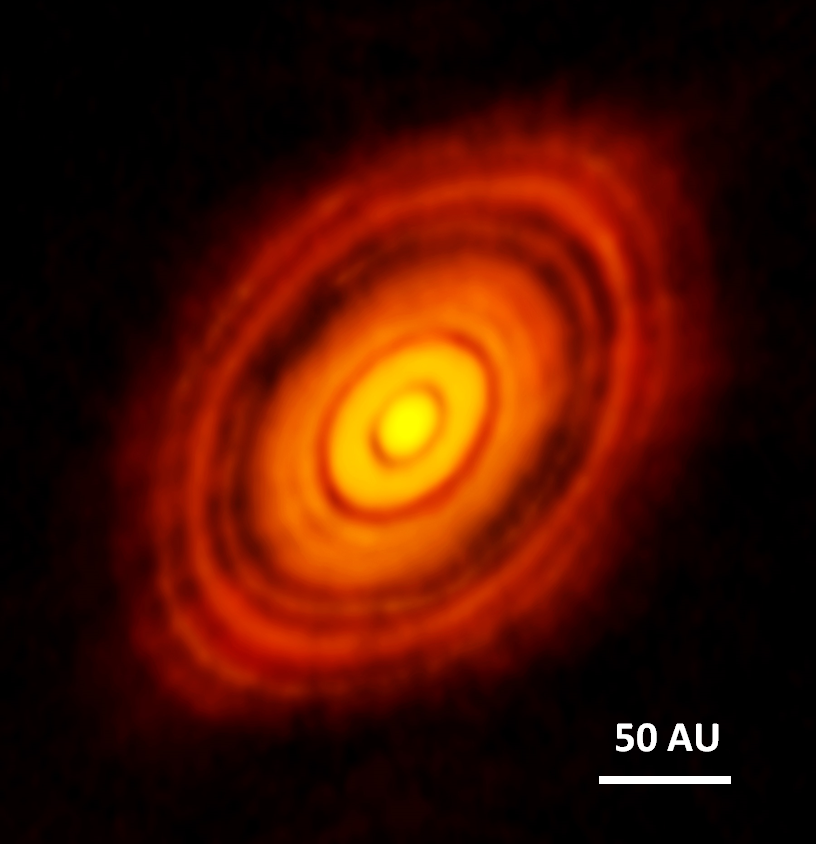}
     \includegraphics[scale=0.45, angle=0]{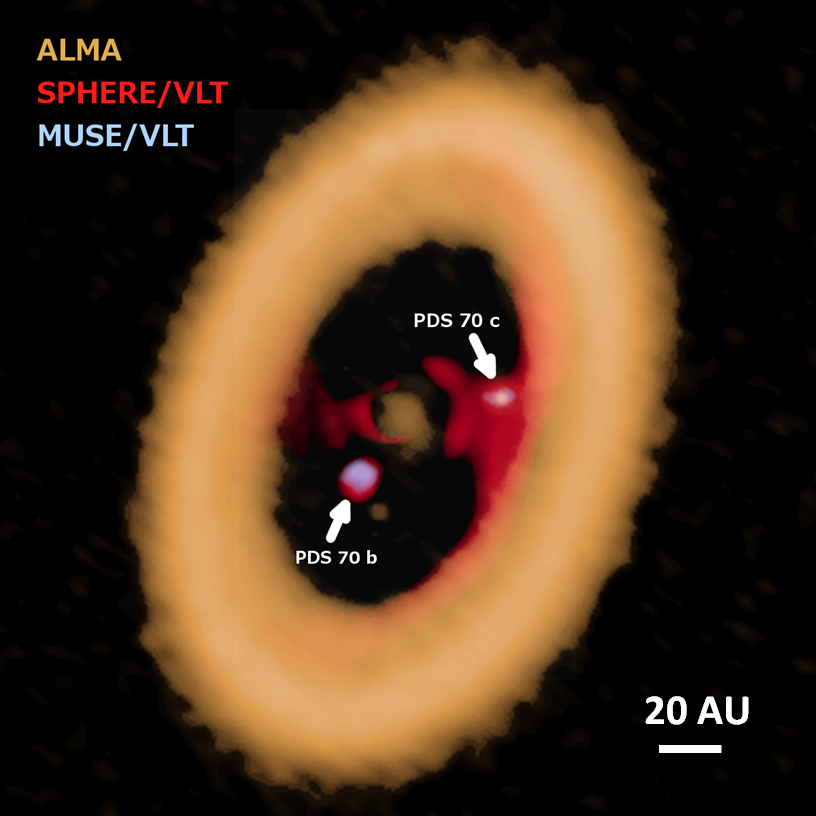}     
   \caption{ Left (a):  protoplanetary disk of HL Tauri with multiple rings and gaps from ALMA dust continuum emission. Credit: ALMA (ESO/NAOJ/NRAO).  Right (b): 
synthetic image of two young planets inside the gap of the disk around PDS $70$. The data are adopted from ALMA,  and  SPHERE and MUSE (The Multi Unit Spectroscopic Explorer) instruments on ESO's Very Large Telescope. Credit: ALMA (ESO/NAOJ/NRAO) A. Isella; ESO. We note that the above two images are not on the same spatial scale.  
   }
   \label{fig:disk}
   \end{figure}

\begin{figure}
   \centering
   \includegraphics[scale=0.95, angle=0]{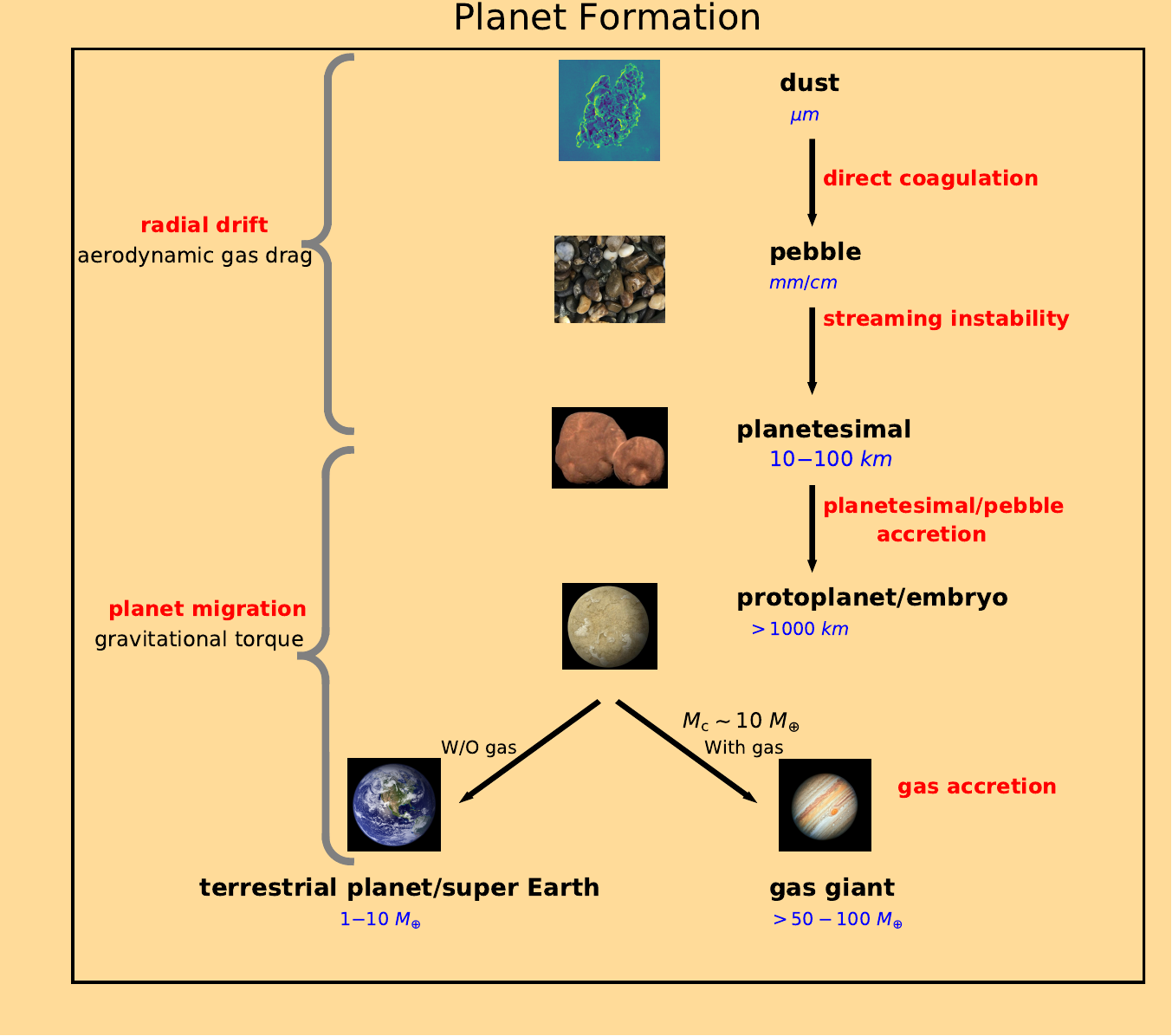}
   \caption{   Planet formation flow chart with characteristic planet bodies including $\upmu$m sized dust grains, mm/cm sized pebbles,  $100$ km sized planetesimals, ${>}1000$ km sized protoplanets and final planets (either gas-deficit terrestrial planets, super-Earth planets or gas-rich giant planets). The physical processes are marked in red text, where the orbital migration is shown in the left while the mass growth is exhibited in the right. 
   }
   \label{fig:body}
   \end{figure}

\subsection{Overview of planet formation}
\label{sec:formation}  
The study of planet formation is a highly multi-scale and  multi-physics  subject. The  size increment of a planetary body varies by more than $17$ orders of magnitude,  from (sub)micron-sized (dust grain) to ${>} 10^{4}$ km (super-Earth/gas giant planet). It results in different physical mechanisms operating at different length scales and in different growth stages. 
 We categorize the planetary bodies  into four characteristic size objects: $ \upmu$m-sized dust grains, millimeter/centimeter (mm/cm)-sized pebbles, $100$-km-sized  planetesimals, and larger than $1000$-km-sized  protoplanets\footnote{ Protoplanets are sometimes also termed  protoplanetary embryos in literature studies.  We do not conceptually distinguish these two words and refer to them as the same  planetary object.}/planets. The final planets are either rocky-dominated terrestrial planets/super-Earths, or gas-dominated giant planets. Chronologically,  the planet formation can be classified into the following three stages: 
from  dust  to pebbles (\Se{dust}),  from pebbles to planetesimals (\Se{streaming}), and 
from planetesimals to protoplanets/planets (\Sess{planetesimal}{PA}{comparison}).

\Fg{body} is a sketch of planet formation with characteristic size bodies and dominant physical processes. Small dust grains coagulate into larger particles in the beginning. Direct sticking is nevertheless stalled for pebbles of roughly mm/cm size \citep{Guttler2010}. Other mechanisms are needed to make the growth of larger bodies proceed.  One leading mechanism is the streaming instability \citep{Youdin2005}, which clusters pebbles and directly collapses into planetesimals by the collective effect of self-gravity. The subsequent growth of planetesimals can proceed by accreting surrounding planetesimals and/or pebbles that drift inward from the outer part of the disk.   When the core mass reaches a critical value (${\sim} 10 \Me$, \citealt{Pollack1996}) and there is still ample disk gas left, the protoplanets can accrete surrounding gas rapidly to  form  massive gas giant planets with a timescale much shorter than the disk lifetime. Otherwise, protoplanets only modestly accrete gas and form low-mass terrestrial planets or super-Earths.

Besides the mass growth, planetary bodies also interact and transfer angular momentum with disk gas, which induce orbital migration.  For instance, solid particles and small planetesimals mainly feel aerodynamic gas drag, and their orbital decay is termed radial drift \citep{Adachi1976,Weidenschilling1977a}.  Meanwhile,  large planetesimals/planets exert gravitational forces  with disk gas, and the corresponding movement is called planet migration \citep{Goldreich1979,Goldreich1980,Lin1986}.

We discuss how planetary bodies grow at each stage in the following sections. Our focus is how  state-of-the-art planet formation models fit into  frontier observational studies.  Due to the length of the paper, we do not go deep into the topics of planet migration and gas accretion in the gas-rich disk phase, and the  long-term dynamical evolution of planetary systems in the gas-free phase. 
The key relevant questions that will be discussed in this paper are listed as follows: 
\begin{itemize} 
\item What  challenges are involved in dust coagulation? What is the characteristic size of the  particles that $\upmu$m-sized dust grains can directly grow into (\Se{dust})?  \\
\item How can  planetesimals form by the streaming instability?  When and where are planetesimals likely to form (\Se{streaming})? \\
\item What are the differences and features  between planetesimal accretion (\Se{planetesimal}) and pebble accretion (\Se{PA})? \\
\item  Which is the dominant accretion channel for  planetesimal/protoplanet growth (\Se{comparison})?\\ 
\end{itemize}

\section{From dust to pebbles}
\label{sec:dust}
In this section we visit the first stage of planet formation: the growth and radial drift of dust grains.  We focus on the comparison between state-of-the-art theoretical and laboratory studies and the latest disk observations.  

\subsection{ Theoretical and laboratory studies }   
\label{sec:pebble_theory}

\subsubsection{ Radial drift of solid particles }   
\label{sec:drift_theory}

Since the gas in protoplanetary disks  is pressure-supported, it rotates around the central star at a velocity $v_{\rm \phi,g} = (1- \eta)v_{\rm K}$, where $\phi$ refers to the azimuthal direction in  cylindrical coordinates,  $v_{\rm K}{\equiv} \sqrt{GM_{\star}/r}$ is the Keplerian velocity at the radial disk distance $r$, $G$ is the gravitational constant and $M_{\star}$ is the mass of the central star.  The headwind prefactor that measures the disk pressure gradient is given by \citep{Nakagawa1986} 
 \begin{equation}
\eta=-\frac{c_{\rm s}^2}{2v_{\rm K}^2} \frac{\partial \ln P_{\rm g}}{\partial \ln r} =\frac{ 1.5- 0.5 k_{\rm T}  - k_{\Sigma}}{2} \left(\frac{H_{\rm g}}{r}\right)^2 ,
\label{eq:eta}
\end{equation}
where $c_{\rm s}$ is the gas sound speed,  $P_{\rm g}$ is the gas pressure, $H_{\rm g}$ is the gas disk scale height, $h_{\rm g}{=}H_{\rm g}/r$ is the disk aspect ratio, and $k_{\Sigma} $ and $k_{\rm T} $ are the gradients of gas surface density and temperature, respectively. The relative velocity between $v_{\rm \phi,g}$ and $v_{\rm K}$ is often referred to as the headwind velocity $\eta v_{\rm K}$. In most regions of the protoplantary disk, the pressure gradient is negative, and gas orbits at a sub-Keplerian velocity. All notations utilized in this paper are listed in Table~\ref{tb:long}.

There are various disk models used for studying young protoplanetary disks,  of which the Minimum Mass Solar Nebula  (MMSN) model is most commonly adopted. The MMSN model  represents the minimum amount of solids necessary to build the Solar System planets \citep{Hayashi1981}, which is given by
 \begin{equation}
\Sigma_{\rm g}= 1700 \left(\frac{ r}{1 \AU} \right)^{-3/2} {\rm  \ g \ cm^{-2}}, \  T_{\rm g} = 280  \left(\frac{ r}{1 \AU} \right)^{-1/2} {\rm \  K}, \ h_{\rm g}  =3.3 \times10^{-2}  \left(\frac{ r}{1 \AU} \right)^{1/4}. 
\label{eq:eta}
\end{equation}
Therefore, at a $1$ AU orbital distance $\eta {= }1.8 \times 10^{-3}$ and the headwind velocity $\eta v_{\rm K} {\simeq} 50 \rm \ m\ s^{-1}$.  Here we choose the MMSN as a canonical disk model.  Other disk models of surface densities  include the minimum mass extrasolar nebula (MMEN) model based on mass extrapolation from short period exoplanets \citep{Chiang2013}, obtained from the  emission of dust disks at (sub)millimeter wavelengths \citep{Andrews2009} or  inferred from the disk accretion rate and viscous theory \citep{Lynden-Bell1974,Hartmann1998}. The disk heating source and dust opacity determine the gas temperature profile. The dominant heating mechanisms include viscous dissipation \citep{ Ruden1986,Garaud2007} and stellar irradiation \citep{Chiang1997,Bell1997,Dullemond2001,Dullemond2004}.       

Unlike gas, a solid particle does not feel the pressure gradient force and tends to move at the Keplerian velocity.  The solid particle experiences a hydrodynamic gas drag  when its velocity deviates from that  of the gas \citep{Whipple1972,Weidenschilling1977a}
\begin{equation}
\vec{F_{\rm drag}}=
 \begin{cases}
{\displaystyle 
 -\frac{4 \pi }{3}  \rho_{\rm g} R^2  v_{\rm th} }  \vec{\Delta v}   \hfill \hspace{1cm}   {\rm [ Epstein \ drag \  law] },   \vspace{ 0.2 cm}\\
  {\displaystyle 
-\frac{ \pi}{2}C_{\rm D} R^2 \rho_{\rm g} \Delta v \vec{\Delta v} }  \hfill \hspace{1cm}   {\rm [ Stokes  \ drag \ law] }, 
\label{eq:Fdrag}
     \end{cases}
\end{equation}
where $R$ is the radius of a spherical particle, $\rho_{\rm g}$ is the gas density, $v_{\rm th}{=} \sqrt{8/\pi}c_{\rm s}$ is the mean thermal velocity of the gas,  $\vec{\Delta v}{=} \vec{v} -\vec{v_{\rm g}}$ is the relative velocity between the particle and gas, and $C_{\rm D}$ is given by 
\begin{equation}
 C_{\rm D}=
 \begin{cases}
  {\displaystyle 
   { 24 R_{\rm e}^{-1}}}      \hfill \hspace{1cm}   \  R_{\rm e} < 1,   \vspace{ 0.1 cm}\\ 
     {\displaystyle 
  { 24 R_{\rm e}^{-0.6}}}       \hfill \hspace{1cm}    \ 1 \leq  R_{\rm e} \leq 800,  \vspace{ 0.1 cm}\\ 
      {\displaystyle 
   { 0.5}   }    \hfill \hspace{1cm}    \  R_{\rm e} >  800.   \vspace{ 0.1 cm} 
     \end{cases}
\label{eq:Cd}
\end{equation}
The drag coefficient $C_{\rm D}$ depends on the particle's Reynolds number $R_{\rm e}{=}2 R \Delta v/\nu_{\rm mol}$, where $\nu_{\rm mol}{=} \lambda_{\rm mfp} v_{\rm  th}/2$ is the  kinematic molecular viscosity, $ \lambda_{\rm mfp} {=} m_{\rm mol}/\sigma_{\rm mol} \rho_{\rm g}$ is the gas mean free path, $m_{\rm mol} {=} \mu m_{\rm H} $ and $\sigma_{\rm mol}{=} 2{\times}10^{-15} \ \rm cm^{-2}$  are the mass and collisional cross section of the gas molecule respectively, and $\mu{=}2.33$ and $m_{\rm H}{=} 1.67\times 10^{-24} \ \rm g$ are the gas mean molecular weight and the hydrogen atom mass respectively. 

 The gas drag acceleration can also be written as $\vec{a_{\rm drag}}{=}-(\vec{v} - \vec{v_{\rm g}})/t_{\rm stop}$, where $t_{\rm stop}$ is the stopping time of the particle, which quantifies how fast the particle adjusts its velocity toward the surrounding gas.  
The stopping time of the particles varies in different regimes. For instance, 
\begin{equation}
 t_{\rm stop}=
 \begin{cases}
  {\displaystyle 
   {  \frac{R\rho_{\bullet}  }{v_{\rm  th}  \rho_{\rm g} }}}    \hfill \hspace{1.5cm}     {\rm when} \  R<9/4 \lambda_{\rm mfp}   \hspace{0.3cm}   {\rm [ Epstein \ regime] },   \vspace{ 0.2 cm}\\ 
      {\displaystyle 
   { \frac{4 R}{9 \lambda_{\rm mfp}} \frac{R\rho_{\bullet}  }{ v_{\rm th}  \rho_{\rm g}  } }   }    \hfill \hspace{1cm}   \   {\rm when}  \  R \geq 9/4 \lambda_{\rm mfp}   \hspace{0.3cm}    {\rm  [Stokes \ regime]}.   \vspace{ 0.1 cm} 
     \end{cases}
\label{eq:tstop}
\end{equation}
 We note that the above expression in the Stokes regime holds when $R_{\rm e}{\lesssim}1$. As a particle's size increases, $F_{\rm drag}$ eventually becomes quadratic in $\Delta v$, and the stopping time is inversely proportional to $\Delta v$.  

The stopping time is also widely expressed in a dimensionless form $\taus{=} t_{\rm stop} \Omega_{\rm K}$, where $\taus$ is termed the Stokes number, and $\Omega_{\rm K}{=}v_{\rm K}/r$ is the Keplerian angular frequency.   Generally, pebbles are considered as mm/cm sized  small rocks.  However, from a hydrodynamical perspective, pebbles are specifically referred to as solid particles with a range of  Stokes number approximately  from $10^{-3}$ to $1$. As will be demonstrated in \se{PA},  particles with such Stokes numbers are marginally coupled to the disk gas, and can be efficiently accreted by larger protoplanetary bodies, such as  planetesimals/planets.   

The radial and azimuthal velocity of the solid particle with respect to the Keplerian motion is given by \cite{Nakagawa1986},  
\begin{equation}
 \begin{cases}
  {\displaystyle 
      v_{\rm r}= -   \frac{2 \taus}{1 + \taus^2} \eta v_{\rm K}  + \frac{1}{1 + \taus^2}v_{\rm r, g}   }       \vspace{ 0.2 cm} \\ 
        {\displaystyle 
      v_{\phi}= -  \frac{1}{1 + \taus^2} \eta v_{\rm K}  +\frac{\taus}{2(1+ \taus^2)} v_{\rm r, g}  },      \\ 
     \end{cases}
\label{eq:peb_velocity}
\end{equation}
where $v_{\rm r, g} $ in the second term on the right side of the equation is the gas radial velocity due to disk accretion, much lower than the headwind velocity $\eta v_{\rm K}$ in the first term.  

From \eq{peb_velocity}, the radial velocity of the solid particle peaks at $\taus {=}1$. For the MMSN, such particles  are roughly meter-sized at $1$ AU and cm-sized at $10$ AU. They are strongly affected by the gas and drift toward the central star within a timescale of approximately $100$ orbits. For pebble-sized particles of $\taus{\approx} 10^{-2}{-}1$ and in the protoplanetary disk regions of $r {\lesssim}{50}$ AU, the radial drift timescale is shorter than the gas disk lifetime (${\sim} 3$ Myr, \citealt{Haisch2001}). Therefore, due to the rapid inward drift, the survival of these high Stokes number particles in protoplanetary disks is a long-standing conundrum in planet formation \citep{Adachi1976,Weidenschilling1977a}.

\subsubsection{ Dust coagulation}
\label{sec:growth_theory}
The primordial solids in protoplanetary disks are  dust grains that originate from the interstellar medium (ISM). These solid particles follow a size distribution $n(R){\propto} R^{-3.5}$, in a range between nanometer and (sub)micron \citep{Mathis1977}. The microphysics of grain growth is not controlled by gravity, but relies on electromagnetic interaction, like the Van der Waals force. Such inter-particle attractive forces bring small dust grains together to form  large aggregates through pairwise collisions.  

The collision outcome depends on the impact velocity between dust particles. 
In the early phase of  low-velocity gentle collision,  (sub)micron-sized dust  stick together to form large aggregates with porous structures.  This is referred to as the ``hit-and-stick'' regime.  As the growth proceeds, their collision velocity increases with the size of the particle. In this intermediate velocity regime, the aggregates result in restructuring through compactification    \citep{Dominik1997,Blum2000}. The above process increases the mass-to-area ratio and thus the Stokes number of the particles, resulting in more energetic collisions. The further mass  growth is terminated by either bouncing or fragmentation due to high velocity, catastrophic impacts \citep{Guttler2010}. This results in particles only growing up to millimeter to centimeter sized pebbles in nominal protoplanetary disk conditions \citep{Zsom2010,Birnstiel2012}.

The sticking and growth patterns of the dust aggregates depend on their material properties. In the inner protoplanetary disk regions of less than a few AUs, silicates are the main constituent of dust grains.  During collisions, these silicate aggregates bounce off or even fragment completely at a threshold velocity of approximately $ 1 \rm  \ ms^{-1}$ \citep{Blum2008}. Meanwhile, for the regions outside the water-ice line, dust grains are dominated by water-ice. The icy or ice-coated aggregates are more porous than silicate aggregates, and thus the bouncing is less evident for them \citep{Wada2011}. In addition, the sticking still occurs at a collision of $10 \rm  \ ms^{-1}$ for the icy aggregates. Due to a higher surface energy and a lower elastic modulus,  these icy aggregates are more sticky compared to silicates \citep{Gundlach2015}. 
Nevertheless, based on the recent laboratory experiments, \cite{Musiolik2019} found that the surface energy of icy aggregates is comparable to that of silicates when the disk temperature is lower than $180$ K. If this is true, it implies that the actual difference in the growth pattern of the above two types of aggregates might be less pronounced than anticipated in the literature (also see \citealt{Gundlach2018} and \citealt{Steinpilz2019}).

Theoretical studies of the global dust coagulation and radial transport have shown that (sub)micron-sized dust grains succeed in growing to mm/cm-sized  pebbles \citep{Ormel2007a,Brauer2008,Zsom2010,Birnstiel2010,Birnstiel2012,Krijt2016,Estrada2016}. The size of pebbles is either regulated by the radial drift  in the outer disk region of $r{\gtrsim} 10$ AU, or limited by bouncing/fragmentation in the inner disk region of $r{<}10$ AU (\eg, Fig. 3 of \cite{Testi2014}).   
Nevertheless, further mass increase by incremental growth becomes problematic, due to the above mentioned  bouncing and fragmentation barriers.

\subsection{ Observational studies }
\label{sec:pebble_obs}

\subsubsection{ Evidence for dust radial drift}
\label{sec:drift_obs}

The most straightforward evidence for radial drift of dust comes from the size comparison between the dust and gas components of the disks. The sizes of dust and gas disks can be separately inferred from millimeter dust continuum emission and the molecular line emission of $\rm CO$ isotopes  \citep{Dutrey1998,Hughes2008}. By these comparisons,  gas disks are generally found to be much smaller than dust disks,  indicating  the radial drift of millimeter dust particles has already taken place at the corresponding ages of the systems \citep{Andrews2012,Ansdell2018,Jin2019,Trapman2020}.

The other evidence is from the commonly observed substructures in protoplanetary disks, such as  cavities, rings and gaps. These features resolve the drift timescale problem; otherwise, pebbles should be depleted in the disk regions of ${r\lesssim}50$ AU within a few Myr (\se{drift_theory}).
The difference in spatial distributions of grains with different sizes is usually used to test the presence of pressure bumps and  a strong indicator of the mobility of dust grains.
The disks with large inner cavities/holes of a few tens of AUs are called `transitional disks' \citep{Calvet2005,Espaillat2014,Owen2016}.  In these disks, different spatial distributions are observed from the emissions of the millimeter continuum,  infrared scattering light and/or molecular lines \citep{DongR2012,vanderMarel2013,vanderMarel2015}. 
One should note that the former emission probes the mm-sized pebbles, while the latter two trace $\upmu$m  grains and gas, respectively. The above phenomena can be explained by the dust filtration effect \citep{Rice2006,ZhuZ2012,Pinilla2012a}, where small grains are tightly coupled to  gas flow, while large pebbles drift toward  and halt at local pressure maxima, resulting in mm-sized dust cavities larger than gas/small grain cavities. Similarly, such variation in the gas and dust components is also resolved in ring-shaped substructures \citep{Isella2016}.

Furthermore, the drift of pebbles also leads to a radial variation of chemical  compositions in  disk gas. For instance,  in the protoplanetary disk the $\rm CO$ molecule condenses into solids outside of the $\rm CO$-ice line while it sublimates into vapor inside.  Since some fraction of $\rm C$ is in refractory materials, the disk gas interior to the $\rm CO$-ice line is expected to  have a $\rm C/H$  lower than the stellar value when pebbles are assumed to be static.  However, when these pebbles continuously drift inwardly and cross the $\rm CO$-ice line, a substantial amount of $\rm C$ is loading interior to the ice line and sublimates into vapor, which naturally increases $\rm C/H$ interior to the $\rm CO$-ice line \citep{Krijt2018}.   
When comparing  the $\rm C/H$  in the stellar photosphere and in the disk gas,  \cite{ZhangK2020} first reported an elevated $\rm C/H$ interior to the $\rm CO$-ice line  in the disk of HD $163296$,  which is $1{-}2$ times higher than the stellar value.  This $\rm C/H$ enrichment is in line with the large-scale radial drift of icy dust particles.

\subsubsection{Constraints on pebble size}
\label{sec:growth_obs}

\textbf{Opacity index}

Let us first look at what we can know from the dust continuum emission of young protoplanetary disks at (sub)millimeter and radio wavelengths.  The observed intensity is $I_{\nu} {=}I_{\nu 0} [1-\exp(-\tau_{\nu})]$, where the subscript $0$ refers to the value at the disk midplane, $\tau_{\nu} {= }\kappa_{\nu}\Sigma_{\rm d}$ is the optical depth from the disk midplane to the surface layer, $\kappa_{\nu}$ is the disk opacity at the corresponding frequency, and $\Sigma_{\rm d}$ and $T_{\rm d}$ are the dust mass and temperature, respectively.  When the disk is optically thin ($\tau{\ll} 1$) at the observed wavelengths, $I_{\nu} \approx \tau I_{\nu 0}  \approx  \kappa_{\nu}\Sigma_{\rm d} I_{\nu 0} $.  Consequently, the observed integrated flux  $F_{\nu} \propto \kappa_{\nu}M_{\rm d} B_{\nu} (T_{\rm d})$ and $M_{\rm d}$ is the  dust mass.  At millimeter wavelengths, the Plank function $B_{\nu}$ is expected to approximately follow the Rayleigh-Jeans law, $B_{\nu} \approx 2 k_{\rm B} T_{\rm d} \nu^2/c^2$, where $k_{\rm B}$ is the Boltzmann constant and $c$ is the light speed.  Therefore,    $F_{\nu} \propto \kappa_{\nu} \nu^2 M_{\rm d} T_{\rm d}$. This means that, if $\kappa_{\nu}$ and $T_{\rm d}$ are known, the dust mass can be estimated from the observed flux.

In addition, the spectral index of dust opacity can be obtained from disk observations.  Assuming that the opacity has a power-law dependence on frequency $\kappa_{\nu} \propto \nu^{\beta}$ and $F_{\nu} \propto \nu^{\alpha}$, we have $ F_{\nu} \propto k_{\nu}\nu^2 \propto \nu^{2+ \beta}$.  Since the spectral index $\alpha$ is measured from the spectral energy distribution, the dust opacity index can be calculated accordingly through $\beta = \alpha -2$.  

The commonly accepted evidence for grain growth is based on the spectral index measurements at millimeter wavelengths \citep{Draine2006}. The interpretation is given as follows. Based on the Mie theory calculation, the maximum grain size matters for the dust opacity.  When the maximum grain size is smaller than the observed wavelength (Rayleigh regime), $\kappa_{\nu}$ is independent of grain sizes and $\beta$ remains high.  When the maximum grain size is larger than the observed wavelength (geometric regime), $\kappa_{\nu}$ decreases with the grain size and $\beta$ drops to a lower value close to $0{-}1$ (Fig. 3 of \citealt{Ricci2010a}). As a result,  grains with a maximum size ${\gtrsim} 1$ mm naturally result in a less than unity spectral index at millimeter wavelengths. In other words, the value of $\beta$ can principally reveal the size of the largest grain in disks. 
In a realistic situation, $\beta$ is also affected by the size distribution, composition and porosity of the dust aggregates. Nonetheless, these dependence-induced uncertainties are generally smaller compared to that due to the maximum grain size (Figure  4 of \citealt{Testi2014}).

It is worth mentioning  that the opacity spectral index observed in ISM gives $\beta_{\rm ISM}{\sim} 1.7 $, while the  inferred spectral index of most protoplanetary disks is $\beta_{\rm disk} \lesssim 1$, much smaller than the typical ISM value.  Studies that combines multi-wavelength observations with detailed modelings suggested the ubiquitous presence of grain growth in disks with a variety of ages and around stars with different masses \citep{Natta2004,Ricci2010a,Ricci2014,Miotello2014,Pinilla2017}. Furthermore, the maximum grain size also correlates with the disk radial distance, where generally centimeter sized particles reside in the inner disk regions and millimeter sized grains are present further out \citep{Perez2012,Tazzari2016}. Noticeably, for disks with substructures, the spectral index also varies across the  rings and gaps, with lowest values in the rings and highest values in the gaps,  indicating further grain growth in the high density centric ring regions \citep{HuangJ2018,LiY2019,Long2020}.

.

The above spectral index interpretation is based on two underlying assumptions:  the dust emission is optically thin, and the opacity is dominated by absorption rather than scattering at the observed wavelengths. These two assumptions may also be intrinsically correlated. For instance, if the scattering is the main source of  opacity instead of absorption,  the observed intensity would be $I_{\nu} {\approx} \sqrt{1- w_{\nu}} \tau_{\nu} I_{\nu 0}$ \citep{ZhuZ2019}, where  $w_{\nu}$ is the single-scattering albedo. The above  formula reduces to  the  previously mentioned $I_{\nu}\approx \tau I_{\nu 0}$ where the scattering is negligible compared to the absorption ($w_{\nu} {\to} 0$).   This means that the scattering causes disks to look cooler than they actually are (\eg, Figure 9 of \citealt{Birnstiel2018}). In this respect, the disk mass and  optical depth are likely to be underestimated when the scattering is ignored in literature studies \citep{ZhuZ2019,LiuH2019,Ballering2019}. If the disks are indeed very massive and optically thick even in millimeter wavelengths, then the above approach for the grain size estimation is not valid anymore.  Recently, \cite{CarrascoGonzalez2019} considered both scattering and absorption in dust opacity and neglected any underlying assumption on the optical depth for the study of the  HL Tau disk. They still found that the grains have grown to millimeter size. Importantly, similar treatments with careful assumptions should be applied to other disks  for more realistic estimates of the grain size as well.

 \textbf{Polarization}

Several studies attempted to explain the disk polarized emission by dust scattering  \citep{Kataoka2016,YangHF2016}, although other interpretations such as grain alignment with the disk magnetic field could still be relevant. If the dust scattering is indeed the dominant mechanism, the maximum grain sizes can also be constrained by the observed  polarization degree. For the same system, the HL Tau disk,  \cite{Kataoka2017} reported a maximum grain size of $100 \rm  \ \upmu m$,  one order of magnitude smaller than the size estimated from  \cite{CarrascoGonzalez2019}.  Despite these polarimetric measurements being only applied to a small number of disks, the inferred size is considerably lower than that obtained from opacity index measurements \citep{Hull2018,Bacciotti2018,Ohashi2020}. 

The above polarization analysis overall supports grain growth. Nevertheless, it is still unclear whether the discrepancy in grain size estimation between these two  interpretations is because of the existence of multi-specie dust grains, or due to the limitations and degeneracies in methodologies themselves.  Future studies are warranted to make further claims.

 \textbf{Meteorites  in the Solar System}\\
There is evidence of grain growth in our Solar System.
For instance, the calcium-aluminium-rich inclusions (CAIs) are sub-mm to cm-sized grains   
identified in the most primitive meteorites.  These refractory inclusions are thought to be the earliest solids condensed from the young nebula that form the Solar System. In cosmochemistry, the $\rm Pb$–$\rm Pb$ isotopic  dating shows that CAIs formed $4.567$ Gyr ago \citep{Amelin2010,Connelly2012}. The size of CAIs supports  the hypothesis of coagulation-driven growth of condensates.  In addition, chondrules are igneous-textured spherules dominating in chondrites. The typical size of chondrules is $0.1$ to $1$ mm       \citep{Friedrich2015,Ebel2016,SimonJ2018}. Although there are still discrepancies between $\rm Pb$-$\rm Pb$ ages and $\rm Al$-$\rm Mg$ ages (see \citealt{Kruijer2020}), chronological studies indicated that a small fraction of chondrules might be formed as early as the CAIs, while the majority formed $2{-}4$ Myr after the formation of CAIs     \citep{Amelin2002,Kleine2009,Villeneuve2009,Connelly2012,Pape2019}.
The presence of chondrules and CAIs in meteorites supports grain growth in the Solar System.

In summary, the current disk observations, in line with theoretical/laboratory studies, together demonstrate  that the first step of planet formation, from dust to pebbles, is robust and ubiquitous  during the protoplanetary disk evolutionary stage.

\section{From pebbles to Planetesimals}
\label{sec:streaming}

As stated in \se{dust}, the direct coagulation fails to produce aggregates much larger than pebble-sized. In this section, we emphasize one  powerful planetesimal formation mechanism termed the streaming instability, which overcomes the above growth barrier by clustering and collapsing dense pebble filaments into planetesimals.
The concept of the streaming instability mechanism and the operating disk conditions are presented in \se{SI_physics} and \se{SI_onset}, respectively. The observational evidence that supports the planetesimal formation by streaming instability is discussed in \se{SI_obs}.

  We also note that there are other alternative scenarios to form planetesimals (see \citealt{Johansen2014}  for a review). For instance,  incremental growth may still proceed when aggregates have very high porous structures without significant compactifications \citep{Suyama2008}.  Such fluffy particles have a higher area-to-mass ratio compared to the compact grains, resulting in a higher collisional rate and a lower  Stokes number.  These highly porous aggregates could still overcome the above growth and radial drift barriers and form planetesimals under certain circumstances    \citep{Okuzumi2012,Kataoka2013,Homma2019}. Besides, turbulent clustering is another mechanism that directly concentrates small dust grains into planetesimals. \citep{Cuzzi2008,Hartlep2020}. Different than the streaming instability, this mechanism requires a prior condition of  underlying turbulence. The optimal size of operating particles is crucially related to the energy cascade models \citep{Cuzzi2010,Pan2011,Hartlep2017,Hartlep2020}.

\subsection{Streaming instability}
\label{sec:SI_physics}

\begin{figure}
   \centering
   \includegraphics[scale=0.5, angle=0]{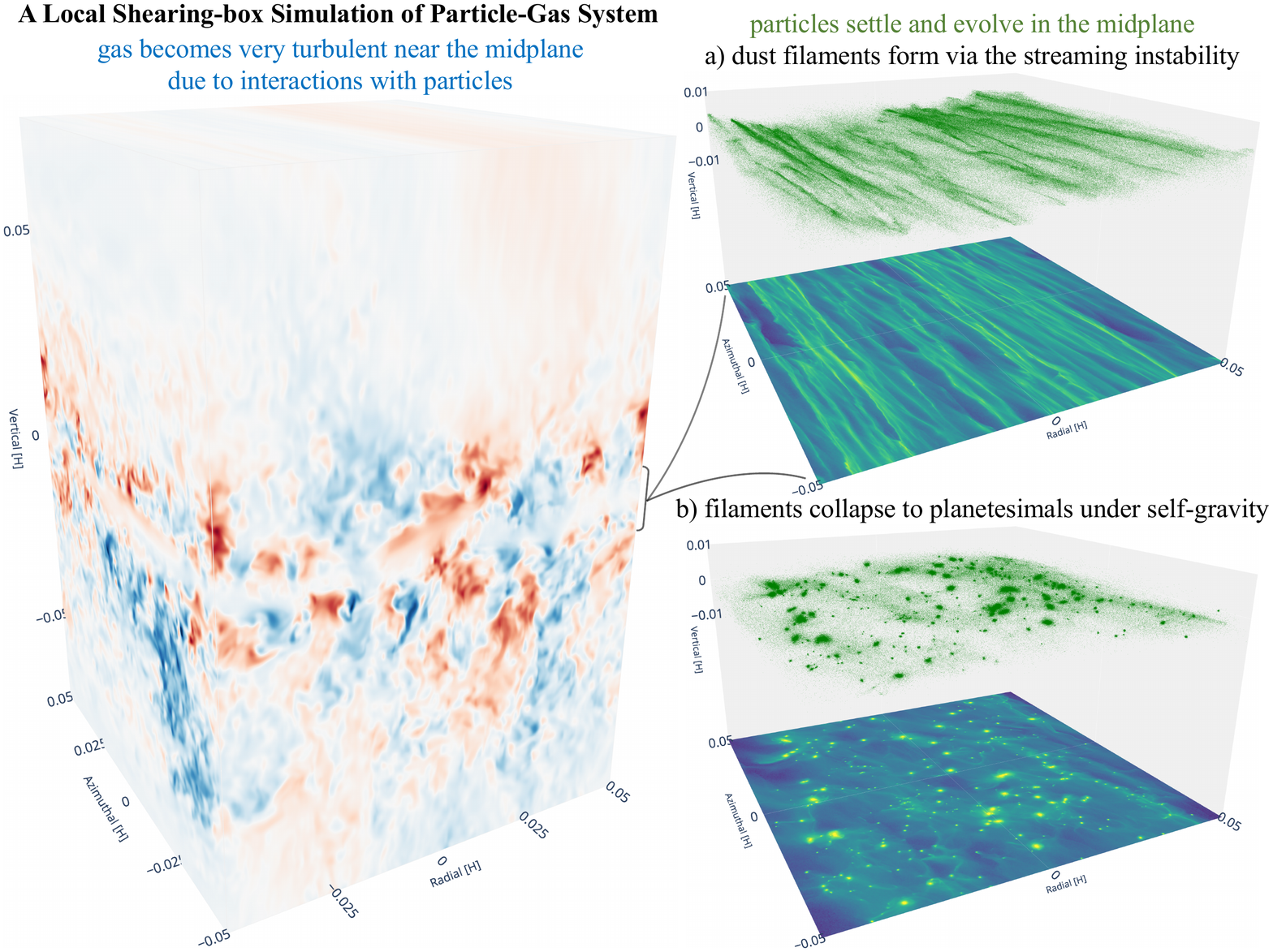}
   \caption{  Illustration of $3$D local shearing-box simulation of the streaming instability for the gas-particle system. Left: the box shows the simulation domain, with each side a fraction of gas scale height. Color maps the vertical gas
momentum, which exhibits high turbulence near the midplane due to interactions between gas and sedimented solid particles.  Right: solid particles (color refers to the surface density) in the midplane (a) form dense filaments due to the streaming instability, and (b) collapse into planetesimals through the collective effect of self-gravity once the solid density exceeds the Roche density. Courtesy of Rixin Li.}
   \label{fig:SI}
   \end{figure}

The streaming instability is applicable for a wide size range of solid particles but most efficiently for the particles of $\taus  {\simeq} 0.1{-}1$. We conventionally use the word `pebbles' hereafter to refer  to the solid particles incorporated in the  streaming instability mechanism.   Basically speaking, the streaming instability includes two processes, namely the concentration of pebbles into dense clumping filaments, and gravitational collapse of pebble filaments into planetesimals. The snapshots of the above processes and the spatial distribution of solids in a streaming instability simulation are illustrated in \Fg{SI}. 

We give a qualitative explanation here. Firstly, inward drifting pebbles get concentrated radially due to the solid-to-gas back reaction \citep{Youdin2005}. This is because pebbles feel the gas drag force and lose angular momentum. Similarly, the surrounding gas gains angular momentum from pebbles due to the solid-to-gas back reaction, and thus the gas velocity gets accelerated.
The strength of this reaction force is determined by the volume density ratio between pebbles and gas ($\rho_{\rm peb}/\rho_{\rm g}$). We usually neglect this back reaction since the pebble density is much lower than the gas density in the nominal protoplanetary disk condition.   However, when the pebble density is comparable to the gas density, the back reaction force is non-trivial. In this situation, the pebble density perturbation grows and concentrates pebbles effectively. Since the gas is accelerated towards  the Keplerian velocity, and the relative velocity between gas and pebbles becomes smaller, these pebbles feel weaker gas drag and drift inward more slowly. Hence, the fast drifting pebbles from the outer part of the disk thereby catch  these slower drifting pebbles and form denser filaments. Because this is a positive feedback, once an initial, radial concentration of solids is achieved ($\rho_{\rm peb}{\simeq}\rho_{\rm g}$), the further clumping is self-amplified, and eventually the pebble density grows rapidly in a non-linear manner.  \cite{Youdin2005} proposed the concept of the streaming instability and provided the analytical linear instability solution. The robustness of the pebble clumping was later confirmed in numerical simulations \citep{Youdin2007b,Johansen2007b}.

Secondly, the effect of self-gravity becomes dominant when the pebble density exceeds the Roche density ($\rho_{\rm R} {=} 9\Omega_{\rm K}^2/4\pi G$). In this circumstance, gravitational force overcomes tidal shear, and the pebble filaments gravitationally collapse into $100$-km-sized planetesimals \footnote{ \cite{Gerbig2020} further suggested that the collapse criterion requires gravitational force to overcome  turbulent diffusion on small scales, which also regulates the sizes of forming planetesimals. }. The primordial idea of the gravitational instability was proposed by  \cite{Goldreich1973}.  At that time these authors only considered that dust sediment into a super thin midplane layer to exceed the Roche density. However, such a vertical solid concentration generates  the Kelvin-Helmholtz instability and develops the  vertical velocity stirring, which prevents the subsequent solid enrichment    \citep{Weidenschilling1980,Cuzzi1993}. On the contrary, the streaming instability induces a radial concentration of solid particles. The  gravitational collapse of dense particle filaments by the above streaming effect was numerically verified by \cite{Johansen2007,Johansen2009}. 

We remark a few things here. Strictly speaking, only the first step is relevant to the concept of the instability -- the streaming motion between solids and gas. Unlike other concentration mechanisms, (\eg, \citealt{Cuzzi2008}), one feature of the streaming effect does not require any underlying disk turbulence.  The second step  does not inherently depend on any initiated mechanisms that cluster pebbles. It broadly represents a  pebble collapsing  process that forms planetesimals  aided by the self-gravity. These two steps have been commonly investigated together in literature studies and have been frequently recognized as a unified mechanism termed the streaming instability.

We also point out that this pebble clumping effect also triggers turbulence, even if the flow of the background gas was originally laminar. As numerically investigated by \cite{LiR2018}, the density fluctuation induced by the streaming instability is not sufficient to halt and concentrate pebbles.  The reason for pebble clumping still arises from the previously mentioned streaming motion between pebbles and gas.

It is noteworthy that most of these streaming instability simulations are conducted in a cubic box centered on a local, co-rotating coordinate frame with fixed Keplerian frequency and radial orbital distance (also  called shearing-box simulations, see \Fg{SI}). The length scale of the box is much smaller than the orbital distance, and therefore, the motion of particles in that box is linearized with the Keplerian shear.  
For most  streaming instability simulations, gas fluid is based on a Eulerian grid, and solid particles are treated as Lagrangian superparticles, each representing a swarm of actual pebbles. Such a particle-fluid hybrid approach substantially reduces the computational cost for investigating the non-linear pebble clumping and planetesimal formation processes.

The mass of  planetesimals formed by streaming instability simulations follows a top-heavy mass distribution, which can be roughly  fitted by a power-law plus exponential decay for the intermediate and high mass branch \citep{Johansen2015,Schafer2017,Abod2019}. A turnover mass may  exist in the lower mass branch  \citep{LiR2019}.  The planetesimals have a characteristic size of ${\sim}100$ km when they form at the asteroid belt region \citep{Johansen2015}. The characteristic mass/size increases with disk metallicity, the mass of the central star and radial distance \citep{Johansen2012,Johansen2015,Simon2016}, modestly increases with gas pressure gradient \citep{Abod2019}, and  appears to be independent of Stokes number \citep{Simon2017}.  
 Based on the extrapolation of literature streaming instability simulations,  \cite{Liu2020} derived the characteristic mass of planetesimals as    
 \begin{equation}
 M_{\rm pl} = 5 \times 10^{-5} \left(\frac{Z}{0.02}  \right)^{1/2} \left(\frac{\gamma}{\pi^{-1}}  \right)^{3/2}  \left(\frac{h_{\rm g}}{0.05} \right)^3  \left(\frac{M_{\star}}{ 1  \ M_{\odot}}  \right)  \ M_{\oplus},
 \label{eq:pl_mass}
  \end{equation}
where  $Z$ is the local disk metallicity and $\gamma {=} 4 \pi G \rho_{\rm g}/\Omega_{\rm K}^2$ is a self-gravity parameter \footnote{We note that  $\gamma$ can be related with the Toomre $Q_{\rm T}$ parameter by $Q_{\rm T} {=}   \sqrt{8/\pi} / \gamma$. Thus, $\gamma {= }0.034$ is equivalent to  $Q_{\rm T} {=} 47$.}, related with gas density, stellar mass and radial distance.  Adopting the MMSN model, we obtain $\gamma {=}0.034$ and the resultant planetesimal is $10^{-6} \Me$ in mass  ($100$ km in radius) at $r{=}2.5$ AU around a solar-mass star.  Based on \eq{pl_mass}, we expect that smaller planetesimals form at shorter orbital distances and around lower-mass stars.

Most of the streaming instability  studies simply considered a laminar  background gas, despite that the protoplanetary disk should be turbulent in nature \citep{Lyra2019}. How the streaming instability operates in a realistic turbulent condition has not been fully understood.
First of all, disk turbulence induces stochastic motion and density fluctuation of the gas. Overdense pressure bumps can be produced at the region in which particles of $\tau_{\rm s} {\sim}1$ are efficiently trapped. This type of solid concentration that facilitates planetesimal formation  is obtained in disks where the source of turbulence is either the  magnetorotational instability (MRI, \citealt{Johansen2007}), or the vertical shear instability (VSI, \citealt{Schafer2020}), by which angular velocity depends on the disk vertical distance \citep{Nelson2013,Stoll2014,LinM2015,Flock2017}.   When considering the non-ideal magnetohydrodynamical (MHD) effects, \cite{YangCC2018} found that dust diffusion is weak in the radial direction and  strong clumping can still occur in the dead zone region where the MRI is inactive because of low ionization fraction  \citep{Gammie1996}.
On the other hand, the turbulent diffusivity acts to suppress sedimentation and concentration of particles, and therefore, the birth rate of planetesimals can be slower or even quenched when the disk turbulence increases \citep{Gole2020}.  These numerical studies have only been explored  in a very narrow range of parameter spaces ($\taus$ and $\alphat$) with various disk turbulence mechanisms. 

In order to quantify the role of turbulence, \cite{Umurhan2020} and \cite{Chen2020} have conducted linear stability analyses for the motion equations of gas and solid particles by including  additional viscous forcing terms. They also found that  the growth rate gets reduced even when the disk is moderately  turbulent.  
Nevertheless, the adopted simplified isotropic, $\alpha$ prescription cannot fully mimic the realistic non-linear pattern of the disk turbulence. For instance,  the aforementioned turbulent-induced zonal flows and pressure bumps seem to promote dust concentration and subsequent planetesimal formation \citep{Johansen2007,Schafer2020}, which is not captured in these theoretical  analyses. 
On the whole, it is still premature to reach definitive conclusions yet, and future studies are required for building a unified and consistent picture of this topic.

\subsection{How, where and when the streaming instability occurs}
\label{sec:SI_onset}

In order to trigger the streaming instability, the volume density of solids needs to be enhanced comparable to that of  gas, $\rho_{\rm peb}{\simeq}\rho_{\rm g}$ \citep{Youdin2005}.  The onset criterion can also be expressed in terms of the surface density ratio, \ie, the metallicity $Z {=} \Sigma_{\rm peb}/\Sigma_{\rm g}$. The  dust to gas  mass ratio is measured to be $0.01$ in the ISM \citep{Bohlin1978}, while the canonical value of the solar metallicity is $0.014$ \citep{Asplund2009}.  Numerical  studies reported that a super-solar metallicity (${\gtrsim}2\%{-5}\%$) is required for triggering the streaming instability \citep{Johansen2009,Carrera2015,YangCC2017}. Such a threshold metallicity also depends on the disk and pebble properties. The streaming instability is easier to be triggered when the disk metallicity is higher \citep{Johansen2009}, the strength of the disk pressure gradient is lower  \citep{Bai2010}, and/or the Stokes number of pebbles is higher towards unity \citep{Carrera2015}.

Nonetheless, unless some other mechanisms can operate in the first place to enhance the pebble density, the disk with solar metallicity can hardly form planetesimals by the streaming instability. Then the question is how the solid density can be enriched to satisfy this condition?

We list several scenarios that propose  pebble enrichment at peculiar disk locations. For instance, the formation site can be the water-ice line \citep{Ros2013,Ida2016b,Schoonenberg2017,Drazkowska2017,Hyodo2019}. This is because the water-ice in pebbles sublimates into vapor when these pebbles drift inwardly across the water-ice line ($T_{\rm g}{\simeq}170$ K). First, pebbles are locally piled-up by a ``traffic jam” effect since the outer fast drifting  icy pebbles catch up with the  inner slow drifting silicate grains.  Second, the released water vapor diffuses back to the outside of the ice line and condenses onto the continuously inwardly drifting icy pebbles. This diffusion and re-condensation process  also enhances the local solid density \citep{Stevenson1988,Cuzzi2004}. The former  mechanism generates ``dry'' planetesimals slighter interior to the ice line while the latter one produces ``wet'' planetesimals with  a substantial water fraction  slightly exterior to the ice line.  Similar processes could also be expected at the ice lines of other volatile-rich species, such as $\rm CO$ and $\rm NH_3$. Besides these ice lines, other possible pebble trapping sites can be the edge of the dead zone \citep{Drazkowska2013,Chatterjee2014,Hu2016,Miranda2017}, the vortex generated by hydrodynamical instabilities \citep{Surville2016,HuangP2018}, and the spiral arms in self-gravitating disks \citep{Gibbons2012}.     

Apart from the above mentioned mechanisms that relate with local disk properties, there are other ways of increasing the disk metallicity. For instance,  \cite{Drazkowska2016} showed that this enrichment can occur in the inner sub-AU disk region as a result of the global dust growth and radial drift.   In addition, pebble trapping is thought to be a natural consequence of giant planet formation. 
The massive planet opens a gap \citep{Lin1986} and produces a local pressure maximum in its vicinity \citep{Lambrechts2014b}. 
 Pebbles drift more and more slowly and get concentrated on their way towards  this local pressure maximum. When these pebbles reach the threshold metallicity at or close to the gap edge, the streaming instability is triggered to form planetesimals \citep{Eriksson2020}.  
 Moreover, the solid enrichment can be fulfilled in the late disk dispersal phase when stellar photoevaporation dominates.  In this case, pebbles have already  decoupled from gas and sedimented to the disk midplane. The photoevaporating wind blows gas away from the disk surface. Therefore, the solid-to-gas ratio increases globally \citep{Carrera2017}.

To summarize, the planetesimal formation can  either occur locally at peculiar disk locations such as ice lines and pressure bumps, or in a wide range of disk regions when stellar photoevaporation globally depletes the disk gas. 
Where and when the planetesimals form crucially depend on the detailed disk conditions and pebble concentration mechanisms. For instance, the formation location can broadly range from the most inner  sub-AU disk region (dead zone edge) to the outer part of the disk extending to a few tens/hundreds of AUs (spiral arms in self-gravitating disks). The formation time is even more difficult to quantify, which might occur at an early phase when the disk is still self-gravitating ( $t {\lesssim} 0.1{-}1$ Myr), or at a late phase when gas is significantly dispersed ($t{>}3$ Myr). The other time constraint is from the meteorite chronology in our Solar System (see \citealt{Kruijer2020} for a review).   What we learn is that  these parent bodies of meteorites are apparently not all formed at once.  They are more likely to form successively or undergo multiple formation phases over the entire disk lifetime. For instance, the parent bodies of iron meteorites formed within the first Myr \citep{Kruijer2014}, whereas those of chondritic meteorites formed slightly later, ${\sim}2{-}4$ Myr after CAI formation \citep{Villeneuve2009,Sugiura2014,Doyle2015}.

\subsection{Evidence for the streaming  instability} 
\label{sec:SI_obs}
One tentative argument that supports the streaming instability mechanism arises from the optical depth measurements of the rings in young protoplanetary disks from the DSHARP survey. All these dusty rings shown in various systems have similar optical depths of the order of unity \citep{Dullemond2018}.  These observed values can be interpreted as  ongoing planetesimal formation regulated by the streaming instability \citep{Stammler2019}. In their explanation, drifting pebbles are concentrated in the ring where  the streaming instability can be triggered.  The streaming instability converts the pebbles into planetesimals when $\rho_{\rm peb}{>} \rho_{\rm g}$, while it is quenched when $\rho_{\rm peb}{<} \rho_{\rm g}$.  Thus, such a regulated process removes the excess pebbles into planetesimals, maintaining the midplane dust-to-gas ratio to be of order unity.  Since the optical depth correlates with the dust-to-gas ratio,  this  interpretation naturally explains the peculiar optical depths in the observed rings. \footnote{ The large-scale dust clumping  at the edges of the rings are also resolved from hydrodynamic simulations \citep{HuangP2020}.}

Evidence of the streaming instability can also be found from the minor bodies in our Solar System.
 \cite{Morbidelli2009} conducted collisional coagulation simulations and found that in order to reproduce the current size distribution of main-belt asteroids, the primordial planetesimals should be ${\gtrsim}100$ km in size.  
Rather than  incremental growth, this characteristic size and the slopes of size distributions of main-belt asteroids and Kupiter belt objects are more consistent with planetesimals obtained from streaming instability simulations \citep{Johansen2015,Simon2016}.

 The most appealing evidence is from the Kuiper belt objects. Recently, a contact binary named Arrokoth (previously known as Ultima Thule or $2014 \  \rm MU_{69}$) was imaged by the New Horizons spacecraft during its flyby.  Arrokoth, resembling other cold classical Kuiper belt objects, is thought to be well preserved in terms of  the pristine properties since its formation. It  consists of two equal-sized, compositionally homogeneous lobes with a narrow contact neck \citep{Stern2019,Grundy2020,Zhao2020}. Such a peculiar shape with little distortion, and the good alignment of the two lobes strongly indicate that this type of object originated from gentle, low-speed mergers of planetesimals within a gravitationally collapsing clump of pebbles \citep{McKinnon2020}.

The prevalence of equal-sized binaries found in the Kuiper belt  supports that they form by the gravitational collapsing mechanism \citep{Nesvorny2010,Robinson2020}. Most of these binaries are in the cold classical Kuiper belt which have low heliocentric orbital inclinations and eccentricities and thus remain primordial  compared to other populations.   Furthermore, these binaries are observed to have similar colors even though the color distribution of the  binary population has  a large intrinsic scatter    \citep{Benecchi2009,Marsset2020}. This is also expected from the gravitational collapsing, since they form from the same reservoir of solids in the pebble clumps.
In addition,  based on the obliquity measurements of trans-Neptunian binaries, \cite{Grundy2019} found that the prograde binaries are more common than the retrograde ones among tight binaries ($22/26$). Such a binary orientation distribution is consistent with the predictions of the streaming instability simulations \citep{Nesvorny2019}. In contrast, the above properties are difficult to fulfill when the binaries form by sequential coagulation and capture  \citep{Goldreich2002}.
 
To conclude, the streaming instability seems to be the widely-accepted and leading mechanism of planetesimal formation. The success of the streaming instability is not only because the  robustness of the mechanism itself is verified by numerous theoretical/numerical work, but also  many key features of the planetesimal populations generated by the streaming instability are consistent with current observations, both within and beyond the Solar System.  
The streaming instability succeeds in bridging the gap between the pebbles and planetesimals. The growth of planetesimals after formation will be discussed in subsequent sections.

\section{Planetesimal accretion}
\label{sec:planetesimal}

We review the planet formation process from planetesimals to planets from \se{planetesimal} to \se{comparison}. In this section we focus on planetesimal accretion. The accretion cross section and accretion rates in different regimes  are described  in  \ses{cross_section}{runaway_oligarchic} respectively. We further discuss  the underlying physical processes  in \se{plt_mechanism} and summarize the key features  and applications in \ses{plt_feature}{plt_application} respectively.

\subsection{Accretion cross section }
\label{sec:cross_section}
The Hill radius of a planetary body orbiting a central star is defined as
\begin{equation} 
R_{\rm H} = \left( \frac{M_{\rm p}}{3M_{\star}}  \right)^{1/3} a,
 \end{equation} 
 where $M_{\rm p}$ and $a$ are the mass and semimajor axis of the body respectively. The Hill velocity is $v_{\rm H} {=}R_{\rm H} \Omega_{\rm K}$.  Within the Hill sphere, the planetary body's gravitational force is more important than that of the star. 

We consider planetesimal accretion in the case of a few massive protoplanetary embryos embedded in a swarm of less massive planetesimals. Hereafter we call these two populations the large and small bodies, and their masses are expressed as $M$ and $m$, respectively. Only the gravitational force operates during their encounters.  The collisional (accretion) cross section of the large body can be expressed as  
\begin{equation}
\sigma{=}\pi R_{\rm M}^2 \left(1 + \frac{v_{\rm esc}^2}{ {\delta v}^2} \right) = \pi R_{\rm M}^2 \left(1 + \frac{2 G M }{ R_{\rm M} {\delta v}^2} \right),
\label{eq:cross_section}
\end{equation}
where $R_{\rm M}$ is the physical radius of the large body,  and $v_{\rm esc}{=} \sqrt{2GM/R_{\rm M}}$ and $\delta v$ are the escape velocity of the large body and the relative velocity between the large and small  bodies respectively. The collision is in the gravitational focusing regime when $\delta v{<} v_{\rm esc}$, while it is in the geometric regime  when $\delta v{\geq} v_{\rm esc}$. The gravitational focusing factor is given by $f_{\rm gf} {=} 1 + {v_{\rm esc}}^2/{\delta v}^2$, representing the enhancement of the collisional cross section compared to the physical cross section ($\pi R_{\rm M}^2$). 

One should note that  \Eq{cross_section}  is valid for the two-body approximation  in the dispersion regime  when $\delta v {\geq} v_{\rm H}$. However, when $\delta v {<} v_{\rm H}$,  Keplerian shear dominates the relative velocity, and the three-body interaction including the gravitational force of the central star becomes important \citep{Ida1989,Lissauer1993}.  

\subsection{Growth modes: runaway   vs. oligarchic }
\label{sec:runaway_oligarchic}

The collision rate of the large bodies by accreting small bodies is given by $n_{\rm m} m \sigma \delta v$,  where $n_{\rm m}{=} \Sigma_{m}/m H$, $\Sigma_{m}$  and $H$ are the number density, surface density and vertical height of the small bodies respectively. The planetesimal accretion rate of the large body  can be written as
\begin{equation}
\dot  M_{\rm PlA}  = \Sigma_{\rm m} \Omega_{\rm K} \pi R_{\rm M}^2\left(1 + \frac{2 GM}{ R_{\rm M} \delta v^2} \right) = 
  \begin{cases}
  {\displaystyle 
   {\frac{ 2 \pi   G M \Sigma_{\rm m}  \Omega_{\rm K} R_{\rm M } }{{\delta v}^2} } \  \propto M^{4/3}}      \hfill \hspace{1cm}  \mbox{when} \  \delta v \ll  v_{\rm esc},   \vspace{ 0.2 cm}\\ 
     {\displaystyle 
   { \pi \Sigma_{\rm m}  \Omega_{\rm K}  R_{\rm M}^2 }  \  \ \  \ \propto M^{2/3}}       \hfill \hspace{1cm}   \mbox{when} \  \delta v \lesssim  v_{\rm esc},   \vspace{ 0.1 cm} 
     \end{cases}
     \label{eq:plt_accretion}
\end{equation}
where $M {\propto} R_{\rm M}^3$ and $H {\simeq} 2 \delta v_{\rm z}/\Omega_{\rm K} {\simeq} \delta v/\Omega_{\rm K}$  for an isotropic planetesimal velocity distribution. The growth timescale is expressed  as $t_{\rm grow}= M/(dM/dt)$, where $t_{\rm grow}$ scales with  $M^{-1/3}$ and $M^{1/3}$  in the former and latter cases, respectively. 

 In the early stage of planetesimal accretion (the former case of \eq{plt_accretion}),  $\delta v$ is mainly excited from the mutual interactions among small planetesimals and therefore remains low compared to  $v_{\rm esc}$. Since the growth becomes faster as $M$ increases, this phase is called the runaway growth  \citep{Safronov1972,Greenberg1978, Wetherill1989,Ida1993,Kokubo1996}. In this circumstance, the mass ratio between the large and small bodies increases with time. Nevertheless, small planetesimals still dominate the total masses of whole populations. 
 
 The above runaway phase cannot last forever.  When the velocities of planetesimals are significantly stirred up by the large bodies (the latter case of \eq{plt_accretion}), the accretion cross sections of large bodies are strongly reduced compared to the runaway case. The growth of the big bodies becomes slower as $M$ increases.  The accretion gradually turns into a self-regulated, oligarchic growth \citep{Ida1993,Kokubo1998}, featured by a decreasing mass ratio among adjacent massive bodies.  In this regime the massive bodies still grow faster than the small planetesimals. After the runaway and oligarchic phases, the system evolves into a bimodal population, with large protoplanets and small planetesimals \citep{Kokubo2000,Thommes2003,Rafikov2003,Ida2004a,Ormel2010b}. These protoplanets have a  typical  orbital separation of $10$ mutual Hill radii \citep{Kokubo1998}. 
  
In the late stage when there is no disk gas left, the protoplanets gradually accrete all residual planetesimals. The random velocities of the protoplanets are fully excited to their escape velocities, and their collisional cross sections reduce to the physical surface areas. In this case, the growth eventually becomes  slow, and the system is chaotic in nature \citep{Agnor1999,Chambers1998,Chambers2001,Raymond2004,Kenyon2006,ZhouJ2007a}.    
 
 \subsection{Relevant physical processes}
\label{sec:plt_mechanism}

The relative velocity $\delta v$ is the most important factor that sets the planetesimal growth regimes. It evolves through a combination of four processes: heating from viscous stirring, cooling from gas drag, dynamical friction and inelastic collisions. Here we briefly discuss them. The detailed derivations can be found in   \cite{Goldreich2004}.

Viscous stirring is a dynamical effect in which the velocity dispersion of planetesimals increases through two-body gravitational encounters. During this process, the system gets dynamically excited. The eccentricities and inclinations of planetesimals relax into Rayleigh distributions. The growth of protoplanets slows down with the increase of random velocities of the planetesimals.  
The timescale for viscous stirring of small planetesimals by large bodies is reported by \cite{Ida1993}, 
\begin{equation}
t_{\rm vs} = \frac{v}{dv/dt} =  \frac{{\delta v}^3}{4 \pi G^2 n_{\rm M} M^2 \ln \Lambda},
\label{eq:t_vs}
\end{equation}
where  $\ln \Lambda \simeq 3$ is the Coulomb factor, and $n_{\rm M}$ is the number density of large bodies. As can be seen from \eq{t_vs},  the effect of  viscous stirring increases with the mass of the large body. 

On the other hand, the random velocity of small planetesimals can also be damped by the gas friction force. The eccentricity damping timescale is given by \cite{Adachi1976}, 
\begin{equation}
t_{\rm e, gas} =    \frac{ m}{ C_{\rm D} \pi R_{\rm m}^2 \rho_{\rm g} \delta v/2},
\label{eq:t_edamp}
\end{equation}
where $C_{\rm D} {=}0.5$ is the gas drag coefficient for planetesimal-sized objects in \eq{Cd} and $R_{\rm m}$ is the radius of the small planetesimal. 

Equating  $t_{\rm vs} =  t_{\rm e, gas}$ and solving for the equilibrium eccentricity of small planetesimals, one can obtain
\begin{equation}
e_{\rm m} = \left(   \frac{8 \ln \Lambda m M \Sigma_{\rm M} a  }{ C_{\rm D} \rho_{\rm g} R_{\rm m}^2  M_{\star}^2 } \right)^{1/5}.
\label{eq:e_m}
\end{equation}
Inserting this value into \eq{plt_accretion},  the growth timescale in the oligarchic regime can be expressed as 
\begin{equation}
t_{\rm og} =   \frac{e_{\rm m}^2 a^2 \Omega_{\rm K}}{2 \pi G \Sigma_{\rm m} R_{\rm M}}=   \frac{ a^2\Omega_{\rm K}}{2 \pi G \Sigma_{\rm m} R_{\rm M}} \left(   \frac{8 \ln \Lambda m M \Sigma_{\rm M} a  }{ C_{\rm D} \rho_{\rm g} R_{\rm m}^2  M_{\star}^2 } \right)^{2/5}.
\label{eq:t_og}
\end{equation}
On the other hand, \cite{Ormel2010b} conducted Monte Carlo simulations for the planetesimal growth and found  that the runaway growth timescale can  be alternatively expressed as  
\begin{equation}
t_{\rm rg} =   C_{\rm rg} \frac{R_{\rm m} \rho_{\bullet}}{\Omega_{\rm K}  \Sigma_{\rm m}},
\label{eq:t_rg}
\end{equation}
 where $C_{\rm rg} {\sim} 10$ is a numerical prefactor.  We note that $R_{\rm m}$ and $\Sigma_{\rm m}$ in \eq{t_rg} are the initial size and surface density of the planetesimals, and the growth timescale only depends on the initial configuration.  This is an important feature of the runaway growth, where the planetesimals spend most of the time doubling their initial masses.   
 
Another important damping mechanism is called dynamical friction. It refers to the process of  equipartition of the random energy between large and small bodies. The consequence of dynamical friction is that the random velocity of the body is  proportional to the square root of its mass  ($\delta v \propto M^{-1/2}$), in which small (large) bodies have high (low) random velocities through their mutual gravitational interactions.  Besides, inelastic collisions also damp the random velocity. When two bodies collide with each other, they conserve total angular momentum but  some fraction of the kinetic energy  transfers into  internal heat.

Now let us recall when is the transition between the runaway growth and oligarchic growth.  Strictly speaking, this transition is not determined by $\delta v/ v_{\rm esc}$ but relies on the relative growth rate of the massive bodies, $(dM/dt)/M \propto M^{\gamma}$. The growth is in the runaway phase  when $\gamma {>}0$ , while the growth is in the oligarchic phase when $\gamma {<}0$.   \cite{Ormel2010b} proposed a physical criterion for the above  transition. At the beginning of the accretion, the mass growth is faster than the excitation of random velocities, corresponding to $t_{\rm rg} < t_{\rm vs}$. The transition occurs when $t_{\rm rg} \sim  t_{\rm vs}$.  After that, the stirring is faster than the growth ($t_{\rm rg} > t_{\rm vs}$), and the accretion enters the self-regulated oligarchic regime. 

\subsection{Features}
\label{sec:plt_feature}

We highlight a few important features of planetesimal accretion. First, the runaway accretion does not necessarily mean that the growth is rapid. The concept of  ``runaway'' refers to a relative growth rate, $d (M_1/M_2)/dt>0$ when $M_1{>}M_2$.   The absolute rate can be high or low, depending on the planetesimal surface density $\Sigma_{\rm m}$, the Keplerian orbital frequency $\Omega_{\rm K}$, and the gravitational focusing factor that relates to ${\delta v}$.

Second, the size of planetesimals matters for the growth. As introduced in \se{streaming}, when planetesimals are assumed to form by the streaming instability, they are primordially large (\eg,  $100$-km in size). These planetesimals are well-decoupled from gas and their orbits remain in-situ during the disk lifetime.  In this circumstance, the massive protoplanets only accrete nearby planetesimals within their feeding zones (${\sim}10$ mutual Hill radii). The final masses of protoplanets are only related to local disk properties (\eg, the planetesimal surface density $\Sigma_{\rm m}$). 

However, if the initial sizes of planetesimals are kilometer or smaller, the above accretion paradigm changes.  Planetesimals of smaller sizes undergo non-negligible radial drift during the disk lifetime. The accreting materials are not limited to local planetesimals in the vicinity of the protoplanets  anymore. Therefore, the planets can finally reach higher masses.  On the other hand, for smaller planetesimals, their random velocities remain lower due to the weaker viscous stirring and stronger gas damping (\eqs{t_vs}{t_edamp}). As a result, the growth of ${<}1$-km-sized boulders is faster compared to the case of $100$-km-sized planetesimals \citep{Coleman2016}. Planet growth has several advantages when the initial size of planetesimals is small. However, the major issue is whether the disk can form planetesimals with dominant size of ${\lessapprox }1$ km.  

Third,  planetesimal accretion also depends on their disk locations.  As shown in \eq{plt_accretion}, when the planetesimals are further out, the accretion is slower due to a longer orbital time. In addition, the encounters tend to result in ejections rather than collisions when the orbital distance is larger.  
 The outcome of the planet-planet encounters  can be quantified as \citep{Goldreich2004,Ida2004a}
 \begin{equation}
 \Phi^2 = \frac{v_{\rm esc}^2 }{2 v_{\rm K}^2} = \left( \frac{M_{\rm}}{M_{\star}} \right) \left(\frac{a_{\rm }}{R_{\rm }} \right),
 \end{equation}
 where $ \Phi$ is the ratio between the escape velocity of the primary body ($v_{\rm esc}$) and the escape velocity of the system ($\sqrt{2}v_{\rm K}$), 
 When $\Phi{\gg}1$, the two-body encounter results in one of them being ejected. On the other hand, the two bodies tend to collide when $\Phi{<}1$.

In short,  planetesimal accretion is faster when the surface density of planetesimals is higher,  the orbits of planetesimals are closer in, and/or their initial sizes are smaller.  

\subsection{Applications}
\label{sec:plt_application}
{\bf Solar System}

The formation timescales of terrestrial planets can be measured by using  radioactive decay of short-lived isotopes, among which the hafnium-tungsten ($\rm Hf{-}W$) isotope is widely adopted for radiometric dating.  This is not only due to its applicable radioactive decay half-life time of $9$ Myr, but is also related to the chemical properties of  these two elements: $\rm Hf$ is lithophile (``rock loving'') and $\rm W$ is siderophile (``iron loving'').  $\rm W$ preferentially settles into the metal core and $\rm Hf$ remains in the silicate mantle before the protoplanet becomes massive enough to segregate. 
For instance, if the core forms early (early mantle-core segregation), the measured $\rm W$ abundance in the mantle would be high. On the other hand, if the core forms late, $\rm W$ would be high in the core, leaving the $\rm W$-deficient mantle. 
Therefore, measuring the $\rm Hf{/}W$ ratio in the current Earth mantle can be utilized to constrain its core formation and differentiation.  Such isotope analyses indicate that core formation of Earth occurred $30{-}100$ Myr after formation of the Solar System \citep{Kleine2002,Yin2002, Jacobsen2005,Touboul2007,Kleine2009,Rudge2010}. 
Similarly, based on Martian meteorites, the  formation time of Mars is inferred to be within a few Myr \citep{Kleine2004,Foley2005,Dauphas2011}, comparable to the lifetime of the proto-solar nebula.

The final accretion stage of Solar System terrestrial planets, or called late giant-impact stage,  is typically modeled by a numerical N-body approach (see \cite{Izidoro2018} for a summary of numerical methods).
 Before this stage, protoplanetary embryos of lunar to Martian masses were already formed in the terrestrial planet forming region by accreting planetesimals (see \se{runaway_oligarchic}). The random velocities of the embryos will be stirred without efficient damping, and planet-planet collisions  frequently occur after  dispersal of the disk gas. On the other hand, giant planets have already grown to their current masses. The influence of gas giants (in particular Jupiter and Saturn) is crucial for sculpting the architecture and water delivery of the asteroid belt  and the inner terrestrial planets \citep{Raymond2017,Zheng2017}.  Numerous numerical simulations have attempted to reproduce the terrestrial planets,  both in terms of the dynamical properties and geochemical accretion timescale  \citep{Chambers2001,Raymond2004,
Raymond2006,Raymond2009,OBrien2006,
Thommes2008,Morishima2010,
Jacobson2014}. The above models are either assumed to have giant planets on their current positions with nearly circular or slightly eccentric orbits.  
However, these models still have  shortcomings.   For instance, the planets emerging out of the Mars forming region have masses comparable to Earth and Venus.  In other words,  the Mars analogs are difficult to produce in the above numerical simulations, known as the ``small Mars problem".  \cite{Hansen2009} first proposed that the proper Mars-size objects can be yielded when the planetesimal disk is truncated at around $1$ AU.    As a  further step, \cite{Walsh2011} proposed that such a  truncation can be physically caused by the inward-then-outward migration of Jupiter and Saturn through planet-disk interactions, which is known as the Grand Tack model. The key ingredient of this  model, the migration of two giant planets, is consistent with our current understanding of the disk migration theory \citep{Kley2012,Baruteau2014} and has been validated by hydrodynamic simulations  \citep{Masset2001, Morbidelli2007,Pierens2008a,ZhangH2010}. 
The appealing point of the Grand Tack model is that it satisfactorily reproduces key observational features of the inner Solar System, such as the orbital and mass distributions of the terrestrial planets,  mass depletion and the chemical compositions of planetesimals in the asteroid belt, and water content of  Earth \citep{OBrien2014}. On the other hand,  \cite{Clement2018} proposed that Mars' growth can also be stunted if the giant planet instability occurs relatively early during the terrestrial planet formation (\eg, ${\lesssim}10$ Myr after the gas disk dispersal).   

It is also worth mentioning that  most of the current N-body simulations only consider  perfect mergers when two bodies collide with each other. However, the random velocities of planetesimals would be excited when the masses of the protoplanets increase. The collisional outcome sensitively relates to the masses and velocities of the colliding bodies, which could be catastrophic disruption, grinding or fragmentation \citep{Leinhardt2012, Agnor2004,Genda2012,LiuSF2015}. Dedicated N-body simulations including realistic collision recipes  showed that the final masses and numbers of surviving planets are comparable to the case when only perfect mergers are considered, but differences still exist in planet spins, eccentricities, core mass fractions and formation timescales \citep{Kokubo2010,Chambers2013,Carter2015, Clement2019}.

{\bf Exoplanetary systems}

The observed exoplanets are more diverse than Solar System planets. One important question is how to form different planet populations, such as gas giants and super-Earths. The growth of giant planets first requires the assembly of massive cores (${\sim}10  \Me$) that can initiate rapid gas accretion \citep{Pollack1996,Ikoma2000,Hubickyj2005,Movshovitz2010} before the dispersal of the disk gas.
For a nominal disk model, the planetesimal isolation mass at a distance of a few AU  (the formation zone of Jupiter and Saturn) is generally lower than this critical mass \citep{Ida2004a}. Therefore, giant planet formation was thought to be challenging, unless the disks are extremely massive.

 Protoplanets approaching such isolation masses can undergo substantial orbital migration \footnote{It worth pointing out that although the migration theory was first proposed in the late  $1970$s \citep{Lin1979,Goldreich1979,Goldreich1980},  it was not considered in the classical formation models of the Solar System planets \citep{Lissauer1993}.  The disk migration theory attracted attention until close-in exoplanets were subsequently discovered \citep{Lin1996}.}. These protoplanets migrate towards and get trapped into MMRs at some zero-torque locations (or called trapping locations, \citealt{Lyra2010,Horn2012,Kretke2012}). Depending on the number of planets and the gas disk mass, the resonant configurations can be disrupted with rapid migration, resulting in frequent planet-planet collisions even in the gas-rich disk phase \citep{Hellary2012,Pierens2013,ZhangXJ2014,Liu2015}. In this circumstance, massive cores can be  attained early enough and subsequently grow into gas giants.  This interpretation prefers that the cores of gas giants only form early when the disk is massive, \eg, the disk accretion rate is higher than ${\sim}10^{-7}\Msyr$. Comparing that with the typical observed disk accretion rates and the $\dot M_{\rm g}{-}M_{\star}$ dependence \citep{Hartmann1998,Natta2006,Manara2016}, it  explains why only a minor fraction of stars harbor gas giant planets  \citep{Liu2015} and why gas giant fraction increases with the mass of the central star \citep{Liu2016}. 
On the other hand, if planet-planet collisions do not occur early enough, planets remain  low-mass and they evolve into super-Earth systems with compact, (near) resonant configurations  \citep{Terquem2007,Ogihara2009,Cossou2014,Ogihara2015,Izidoro2017}.  Numerous works also explored the detailed resonant trapping and stability of multiple super-Earth systems both numerically \citep{WangS2012,WangS2014,Sun2017,Pan2020} and analytically \citep{Quillen2011,Hadden2018,Petit2020}.

When examining the giant planets in the Solar System and based on the results of the Juno mission, the core of Jupiter is inferred to be diluted with an extended layer of heavy elements \citep{Wahl2017,Helled2017}. The growth of  proto-Jupiter has a strong influence on nearby planetesimals and protoplanets \citep{ZhouJ2007b,Ida2013}.
Such a diluted internal structure is likely to result from the giant impacts among protoplanets during their assembling phase \citep{LiuSF2019} .

The other possible explanation of super-Earths is that they only form late and in-situ when disk gas dissipates significantly. In that case, gas damping is weak and the super-Earth cores can grow quickly through the mergers of protoplanets \citep{Lee2016}. This could also explain why  observed massive super-Earths (${\sim}10 \Me$) do not undergo runaway gas accretion to become  gas giants \citep{Lee2014}. The  exact compositions and orbital properties of the formed super-Earths are determined by a combined effects  of gas damping and solid accretion \citep{Dawson2016}, and potential further out giant planets \citep{Ji2011,Jin2011}. On the other hand, recent hydrodynamic studies found that because the gas recycles efficiently between the planetary envelopes  and surrounding disk, the rapid gas accretion onto  super-Earths can be actually protracted  \citep{Ormel2015,Cimerman2017,Lambrechts2017,Kuwahara2019}. This also provides an interpretation for the ubiquitous presence of super-Earths but not gas giant planets.

The population synthesis model is an ideal tool to explore the influence of key physical processes on planet formation and evolution.  
In this approach, different physical processes are simplified into specialized recipes and combined into a unified deterministic model. By Monte Carlo sampling  the initial conditions over appropriate distributions, the synthetic planetary populations can be  generated and thus compared to the observed exoplanet sample in a statistical manner (see \cite{Benz2014} for a review). \cite{Ida2004a} first applied such a calculation to  investigate  planet formation around Sun-like stars, and further extended their study to systems around stars of various masses and metallicities \citep{Ida2004b,Ida2005}. The predicted correlation between gas giant planets and their stellar hosts exhibited  good agreements with RV measurements  \citep{Fischer2005, Johnson2007}. Sophisticated population synthesis models based on planetesimal accretion are further developed to make testable observational comparisons and to study how forming planets are related to the initial disk and stellar properties \citep{Mordasini2009, Mordasini2012a,Mordasini2012b,Ida2013,Jin2014,Coleman2014,Coleman2016,Alibert2017a,Mulders2019,Miguel2020}.

{\bf Stellar Binary System}

The previously mentioned studies focus on  planet formation around single stars. However, nearly half of Sun-like stars are in binaries \citep{Duquennoy1991,Raghavan2010}, and this fraction is even higher for higher-mass stars \citep{Kouwenhoven2007}. Thus, studying how planets form  in stellar binary systems is of crucial importance. Up to now, more than $150$ exoplanets have been discovered in stellar binary systems, including  both S-type (satellite-like orbits around one of the single stars) and P-type (planet-like orbits around binary stars). P-type planets are also referred to as circumbinary planets.

Planets are less common in binary systems compared to single hosts \citep{WangJ2014b,WangJ2014a}. The S-type planets  have not yet been found in binaries with a period less than $1000$ days. Close binary companions play a destructive role in forming  S-type planets \citep{Thebault2015}. First, the protoplanetary disk would be tidally truncated by the secondary companion, reducing the disk mass and lifetime \citep{Artymowicz1994,Miranda2015}. Second,   secular perturbations induced by the companion excites high relative velocities among planetesimals, leading to catastrophic collisions (\citealt{Heppenheimer1978,Thebault2006},  but also see \citealt{Xie2010a}). Third, hydrodynamic simulations showed that the dynamics of disk gas  in the binary system is much more complicated than typical assumed static, axisymmetric configurations in theoretical analyses \citep{Paardekooper2008,Kley2008,Marzari2009,Muller2012}. All above effects are detrimental for the growth of planetesimals. A noteworthy point is that, disruptive collisions among planetesimals would produce a reservoir of dust debris. The sweeping of the dust debris  can nevertheless boost further mass growth of surviving, leftover planetesimals \citep{Paardekooper2010,Xie2010b}.

Although in-situ formation of S-type planets  in close binaries seems to be very challenging, it is comparatively easy to form P-type planets.
Disk-driven migration needs to be considered when the planets grow massive \citep{Pierens2008b,Kley2015}. The subsequent evolution of planetary systems includes the MMR capture, excitation of eccentricities along with the gas disk dissipation and planet–planet scattering. \cite{Gong2018} suggested that S-type planets can form through planet-planet scattering from P-type planets and then tidally capture in various binary configurations. A smaller eccentricity or a lower mass ratio of the binary leads to a higher capture probability up to $10 \%$ and produces S-type planets with retrograde orbits, consistent with the result of the two unequal-mass planet ejection model \citep{Gong2017}.

Another important topic is the orbital configuration of circumbinary planets (P-type orbits).  More than $20$ circumbinary planets  have been detected so far \citep{Schwarz2016}.  For those discovered by the Kepler mission, the planets are inferred to be inclined by less than a few degrees relative to the binary plane \citep{Kostov2014}. However, the above coplanarity might be caused  by the observational bias: planets orbiting in the binary plane are easier to detect. On the other hand,  circumbinary disks with high inclinations have been subsequently discovered \citep{Kennedy2012,Brinch2016,Czekala2019}, even on polar orbits \citep{Kennedy2019}.  Such misaligned disks may strongly indicate the existence of high-inclination circumbinary planets. In addition, since most stars  originate from star clusters and stellar associations \citep{Lada2003},   the planetary orbits can be influenced by  perturbations from stellar fly-bys. This idea has been explored mainly for planetary systems around Sun-like single stellar hosts, and the corresponding studies generally found that close encounters from  the passing-by stars are prone to be destructive for planetary systems, reducing their multiplicities with surviving planets on more eccentric and inclined orbits. \citep{Spurzem2009,Malmberg2011,Pfalzner2013,LiuHG2013,Hao2013,LiGJ2015,Zheng2015,Cai2017,LiDH2019}.  Similarly, the cluster environment can also significantly impact the planets in binary systems. \cite{Ma2020} showed that these stellar fly-bys can affect the inclination distribution of circumbinary planets. For instance, for close binary systems originating in open clusters with a spacing of ${>}1$ AU, a few repeated fly-bys can already induce their highly-inclined orbits.

\section{Pebble accretion}
 \label{sec:PA}

In this section, we recapitulate pebble accretion, including the physical mechanism (\se{criterion}), the accretion rates and efficiencies in different regimes (\se{regime}), the properties (\se{feature}) and  applications (\se{application}).     

\subsection{Onset and terminal condition}
 \label{sec:criterion}
In contrast to planetesimal accretion, pebble accretion refers to  pebble-sized (${\sim}$ mm-cm) solid particles  that get accreted by planetary bodies \citep{Ormel2010a,Lambrechts2012}. Since these small pebbles are strongly influenced by the surrounding disk gas,  both gas drag and gravitational force play decisive roles during the pebble-planet encounter (also see  reviews  of \citealt{Johansen2017} and \citealt{Ormel2017}). 
Here we provide a physical picture of pebble accretion based on the order-of-magnitude timescale analysis.  How and in which condition pebble accretion commences are discussed as follows. 

During a pebble-planet encounter, the operation of pebble accretion needs to satisfy the following  two conditions: 
\begin{enumerate}
\item The time a pebble settled onto the planet $t_{\rm set}$ is shorter than the  pebble-planet encounter time $t_{\rm enc}$; otherwise, the pebble cannot be accreted onto the planet.  
\item The stopping time of the pebble $t_{\rm stop}$ is shorter than the  pebble-planet encounter time $t_{\rm enc}$, which means that gas drag matters during pebble-planet interaction; otherwise, the pebble-planet interaction is similar to the planetesimal-planet interaction, where only gravitational force plays a role.    
\end{enumerate}

\begin{figure}
   \centering
   \includegraphics[scale=0.8, angle=0]{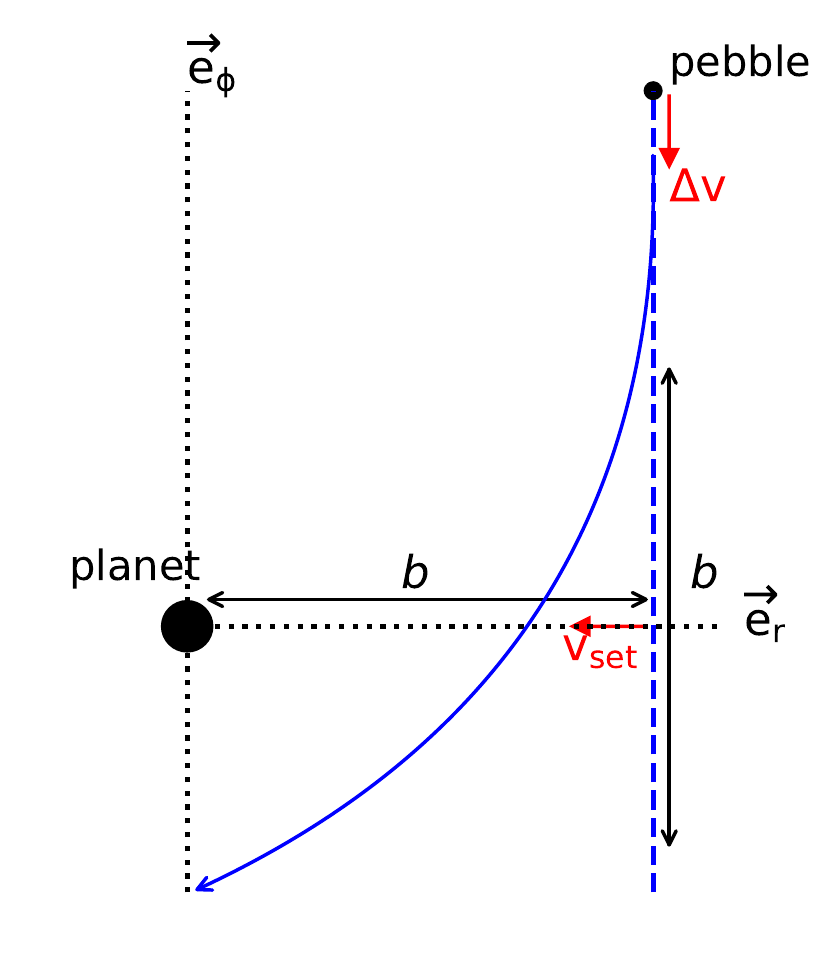}
   \caption{Sketch of the pebble-planet encounter, viewed in a co-moving frame with a central planet. The approach velocity is given by the unperturbed relative velocity between these two bodies ($\Delta v$). The perturbed and unperturbed trajectories are represented in the blue solid and dashed lines respectively with an impact distance parameter $b$. The important timescales are: the stopping time of the pebble $t_{\rm stop}$, the pebble-planet encounter time $t_{\rm enc} = b/\Delta v$, and the settling time $t_{\rm set} = b/v_{\rm set}$, where $v_{\rm set}$ is the sedimentation velocity when the planet’s gravity equals  the gas drag. Figure adopted from \cite{Liu2018}.
   }
   \label{fig:encounter}
   \end{figure}

 An illustration of the pebble-planet encounter is displayed in \Fg{encounter}. When the pebble closely interacts with the planet, the gas drag acceleration  $a_{\rm drag} {=} (v_{\rm peb}- v_{\rm g})/t_{\rm stop}$ is adjusted quasi-statically to balance the gravitational acceleration $a_{\rm g} {=} GM_{\rm p }/b^2$, where $b$ is the impact parameter during the encounter and $t_{\rm stop}$ is the stopping time of the pebble, which quantifies how fast the orbit of the pebble is adapted to the gas due to the gas friction force. The pebble reaches a terminal velocity to sediment onto the planet such that $v_{\rm set} 
{\simeq} GM_{\rm p}t_{\rm stop}/b^2$. 
The settling time and encounter time are given by $t_{\rm set} {= }b/v_{\rm set}$ and $t_{\rm enc} {=} b/\Delta v$ respectively (see \Fg{encounter}), where $\Delta v$ is the unperturbed relative velocity between the planet and pebble. When the above two terms are equal, it gives the largest pebble accretion radius in the settling regime  \citep{Ormel2010a,Lambrechts2012,Ida2016}
 \begin{equation}
b_{\rm set}\simeq \sqrt{\frac{G M_{\rm p} t_{\rm stop}}{\Delta v} }.
\label{eq:PA_radius}
 \end{equation}
As can be seen from the above equation, $b_{\rm set}$ is smaller as $t_{\rm stop}$ becomes shorter. 
 Physically, this is because smaller pebbles with a shorter $t_{\rm stop}$ are more tightly coupled to gas. 
 The trajectories of these pebbles become strongly affected by the planet only when they sediment deeply enough to the planet. The accretion radius is therefore smaller.
 
 Equivalently, criterion (1) can also be derived from the concept of gravitational deflection \citep{Lambrechts2012}.  Pebble accretion occurs when the gravitational deflation time $t_{\rm g} {=} \Delta v / a_{\rm g} =\Delta v / (GM_{\rm p }/b^2) $ is shorter than the pebble stopping time $t_{\rm stop}$.  \Eq{PA_radius} can be obtained by equating $t_{\rm g}$ with $t_{\rm stop}$. 
 
 Criterion (2) breaks down when the unperturbed encounter velocity $\Delta v$ is so high that   $t_{\rm stop} {\geq} t_{\rm enc}$. In this limit, gas drag is too weak to influence the orbit of the pebble during this short timespan.  By equating the above two timescales,  we obtain a threshold velocity  
$v_{\ast}{=}  \sqrt[3]{M_{\rm p} / M_{\star} \tau_{\rm s}} v_{\rm K}$.
 When $\Delta v \ll v_{\ast}$,  pebble accretion is in the settling regime \citep{Ormel2010a} (equivalent to the strongly-coupled regime in  \citealt{Lambrechts2012}). When $\Delta v \gtrsim v_{\ast}$, the accretion enters the inefficient ballistic regime (weakly-coupled regime), where gas drag is unimportant. 
 An exponential decay function is applied to fit such a transition \citep{Ormel2010a,Liu2018}, 
   \begin{equation}
f_{\rm set}= \exp \left[-0.5 \left(\frac{\Delta v}{ v_{\ast}}\right)^2  \right],
\label{eq:fset}
 \end{equation}
 and the accretion radius is expressed as $b_{\rm PA} = b_{\rm set} f_{\rm set}$. 
 
 In order to distinguish from planetesimal accretion, conventionally, pebble accretion refers to the settling regime when the growth is efficiently assisted by the gas drag ($f_{\rm set} \cong 1$).  However, when the mass of the planet/planetesimal is low or $\tau_{\rm s}$ is high, the settling condition is not always satisfied and the accretion can be  inefficient. We can obtain the critical mass for the onset of efficient pebble accretion ($f_{\rm set} \cong 1$) by equating  $v_{\ast}$ with $\Delta v$ (adopted to be the headwind velocity $\eta v_{\rm K}$, see \se{regime}), which gives 
    \begin{equation}
M_{\rm onset}= \taus \eta^{3} M_{\star} = 2.5\times10^{-4} \left(  \frac{\taus}{0.1} \right) \left(  \frac{\eta}{2 \times10^{-3}} \right)^{3} \left(  \frac{M_{\star}}{\Msun} \right) \Me,
\label{eq:onset}
 \end{equation}
corresponding to a $500$ km radius planetesimal \footnote{The above expression is consistent with Eq. 25 of \cite{Visser2016}.}.
 Since the typical planetary bodies we consider are much more massive than $M_{\rm onset}$, for simplicity, the referred pebble accretion hereafter  only corresponds to  the accretion in the settling regime ($b_{\rm PA} = b_{\rm set}$). 
 
The above analysis holds when the planet feedback onto the disk gas is negligible. However,  the  planet perturbation increases with its growing mass. When the planet is massive enough to open a gap and reverse the local gas pressure gradient, the inward drifting pebbles  stop at the local pressure maximum.   In this case, pebble accretion cannot proceed, quenching the core mass growth. Such a process leads to a  `pebble isolation' from the planet, and the onset mass of the planet is called pebble isolation mass \citep{Lambrechts2014b,Bitsch2018,Ataiee2018}.
 For instance, \cite{Bitsch2018} obtained a fitting formula for the pebble isolation mass from hydrodynamical simulations, which reads   
 \begin{equation}
\begin{split}
 M_{\rm iso} = & 25 \left( \frac{h_{\rm g}}{0.05} \right)^3 \left( \frac{M_{\star}}{M_{\odot}} \right) 
 \left[ 0.34 \left(  \frac{ -3}{ {\log}  \alpha_{\rm t}}  \right)^4 + 0.66 \right]\\ 
 & \left[  1- \frac{\partial {\ln} P_{\rm g} / \partial {\ln}r +2.5 }{6}  \right]  M_{\oplus},
 \end{split}
  \label{eq:m_iso}
 \end{equation}
 where $\alpha_{\rm t}$ is the efficiency factor of the turbulent viscosity where $\nu_{\rm t} {=} \alpha_{\rm t} c_{\rm s} H_{\rm g}$ \citep{Shakura1973}.  
 The termination of pebble accretion corresponds to $10{-}20\Me$ super-Earth planets around solar-mass stars and Earth-mass planets around $0.1\Msun$ late M-dwarf stars.

\subsection{Accretion rate and efficiency}
 \label{sec:regime}
As shown in  \eq{PA_radius}, the accretion radius depends on the relative velocity between the pebble and planet $\Delta v$.  For a planet on a circular orbit, $\Delta v$ is a sum of the headwind velocity $\eta v_{\rm K}$ and the Keplerian shear velocity $\Omega_{\rm K}b_{\rm set}$.  The accretion is in the headwind regime (Bondi regime)  when $\Delta v$ is dominated by $\eta v_{\rm K}$, while it is in the shear regime (Hill regime) when $\Delta v$ is dominated by the Keplerian shear. The transition planet mass  between the headwind and shear regimes can be calculated by equating the above two velocities,
   \begin{equation}
M_{\rm hw/sh}= \eta^3 M_{\star}  / \tau_{\rm s} \simeq 0.025 \left( \frac{\eta}{2 \times10^{-3}} \right)^{3} \left(\frac{\taus}{0.1} \right)^{-1} \left(\frac{\Ms}{\Msun} \right) \Me.
\label{eq:Mtran}
 \end{equation}
  When a planet is on an eccentric and inclined orbit,  $\Delta v$ is additionally contributed by the epicyclic motion ($ev_{\rm K}$ and $iv_{\rm K}$) of the planet relative to its Keplerian velocity \citep{Johansen2015,Liu2018,Ormel2018}.      
  
  The pebble mass accretion rate can be expressed as  \citep{Lambrechts2014,Morbidelli2015}
  \begin{equation}
\dot M_{\rm PA}  =
  \begin{cases}
  {\displaystyle 
   2b_{\rm set} \Delta v \Sigma_{\rm peb} \simeq  2 \sqrt{{GM_{\rm p} t_{\rm stop} \Delta v }   } \Sigma_{\rm peb} }
        \hfill \hspace{1cm}  \mbox{[2D]},   \vspace{ 0.2 cm}\\ 
     {\displaystyle 
   b_{\rm set}^2 \Delta v \rho_{\rm peb} \simeq    \frac{GM_{\rm p} t_{\rm stop} \Sigma_{\rm peb}} { \sqrt{2\pi}  H_{\rm peb} }   }   
       \hfill \hspace{1cm}   \mbox{[3D]},   \vspace{ 0.1 cm} 
     \end{cases}
     \label{eq:peb_accretion}
\end{equation}
  where $\Sigma_{\rm peb}{ =} \sqrt{2\pi} H_{\rm peb} \rho_{\rm peb} $. The scale height of the pebble disk is given by \cite{Youdin2007a},
    \begin{equation}
    H_{\rm peb} = \sqrt{ \frac{\delta_{\rm d} }{ \delta_{\rm d} + \tau_{\rm s} }} H_{\rm g},
       \label{eq:H_peb}
\end{equation}
 where $\delta_{\rm d}$ represents the coefficient  of the gas diffusivity, which approximates to the efficiency factor of the turbulent viscosity $\alpha_{\rm t}$ when the disk turbulence is driven by the  MRI \citep{Johansen2005,ZhuZ2015} \footnote{ It is worth noting that the non-ideal MHD effects, such as ambipolar diffusion and Hall Effect, also play important roles in  distributing angular momentum of disk gas. The above processes crucially depend on the disk chemistry and the geometry and strength of the magnetic field \citep{Bai2013,Bai2015,Bai2016,Gressel2015}.}.  In order to distinguish the dominant accretion regime, one can numerically compare $\dot M_{\rm PA, 2D/3D}$ to see which one is higher for given disk and planet parameters.  From a physical perspective, whether the accretion is in $2$D/$3$D is determined by the ratio between the pebble accretion radius $b_{\rm set}$ and the pebble scale height $H_{\rm peb}$ \citep{Morbidelli2015}.  When the pebble accretion radius is larger than the pebble scale height, it is in the $2$D accretion regime; otherwise, it is in the $3$D accretion regime.
  
The disk pebble flux that bypasses the orbit of the planet is $\dot M_{\rm peb} = 2 \pi r \Sigma_{\rm peb} v_{\rm r}$. The pebble accretion efficiency is defined as the probability of pebbles accreted by the planet, $\varepsilon_{\rm PA}  = \dot M_{\rm PA}/ \dot M_{\rm peb}$ \citep{Guillot2014,Lambrechts2014}.
When the radial velocity of the gas  is neglected in $v_{\rm r}$ (\eq{peb_velocity}), the efficiency can be written as
  \begin{equation}
\varepsilon_{\rm PA}  = 
  \begin{cases}
  {\displaystyle 
4\times 10^{-3} \left(\frac{M_{\rm p} }{0.02 \Me} \right)^{1/2} \left( \frac{\tau_{\rm s} }{0.1} \right)^{-1/2}   \left( \frac{\eta}{1.8 \times 10^{-3}} \right)^{-1/2}  \left(\frac{M_{\star}}{\Msun} \right)^{-1/2} }      \hfill \hspace{0.5cm}  \mbox{[2D \ headwind]},   \vspace{ 0.2 cm}\\ 
 {\displaystyle 
4 \times 10^{-3} \left(\frac{M_{\rm p} }{0.02 \Me} \right)^{2/3} \left( \frac{\tau_{\rm s} }{0.1} \right)^{-1/3}   \left( \frac{\eta}{1.8 \times 10^{-3}} \right)^{-1}  \left(\frac{M_{\star}}{\Msun} \right)^{-2/3}     }      \hfill \hspace{0.5 cm}  \mbox{[2D \ shear]},   \vspace{ 0.2 cm}\\ 
     {\displaystyle 
4\times 10^{-3} \left(\frac{M_{\rm p} }{0.02\Me} \right) \left( \frac{h_{\rm peb} }{3.3 \times 10^{-3}} \right)^{-1}   \left( \frac{\eta}{1.8 \times 10^{-3}} \right)^{-1}  \left(\frac{M_{\star}}{\Msun} \right)^{-1}  }
     \hfill \hspace{1cm}   \mbox{[3D]},   \vspace{ 0.1 cm} 
     \end{cases}
     \label{eq:eff_pebble}
\end{equation}
where $h_{\rm peb}{=} H_{\rm peb}/r$ is the pebble disk aspect ratio.  The full expression of $\varepsilon_{\rm PA}$  including the eccentricity and inclination dependences can be found in \cite{Liu2018} and \cite{Ormel2018}. The pebble accretion efficiency is a crucial quantity in planet formation,  since it corresponds to how efficiently the disk pebble mass can be converted into planet mass.

The pebble accretion timescale is therefore given by
 \begin{equation}
        \begin{multlined}
    t_{\rm PA} = \frac{M_{\rm p }}{\dot M_{\rm peb} \varepsilon_{\rm PA} } \simeq \\
     \begin{cases}
  {\displaystyle 
5\times 10^{4} \left(\frac{M_{\rm p} }{0.02 \Me} \right)^{1/2} \left( \frac{\tau_{\rm s} }{0.1} \right)^{1/2}   \left( \frac{\eta}{1.8 \times 10^{-3}} \right)^{1/2} } \\
{\displaystyle   \left(\frac{\dot M_{\rm peb}}{10^{-4} \rm  \Me \ yr } \right)^{-1} \left(\frac{M_{\star}}{\Msun} \right)^{1/2}  \rm  \ yr}      \hfill \hspace{0.5cm}  \mbox{[2D \ headwind]},   \vspace{ 0.3 cm}\\ 
 {\displaystyle 
5 \times 10^{4} \left(\frac{M_{\rm p} }{0.02 \Me} \right)^{1/3} \left( \frac{\tau_{\rm s} }{0.1} \right)^{1/3}  \left( \frac{\eta}{1.8\times 10^{-3}} \right) } \\
{\displaystyle   \left(\frac{\dot M_{\rm peb}}{10^{-4} \rm  \Me \ yr } \right)^{-1}   \left(\frac{M_{\star}}{\Msun} \right)^{2/3}    \rm  \ yr  }      \hfill \hspace{0.5 cm}  \mbox{[2D \ shear]},   \vspace{ 0.3 cm}\\ 
     {\displaystyle 
5\times 10^{4}  \left( \frac{h_{\rm peb} }{3.3 \times 10^{-3}} \right)   \left( \frac{\eta}{1.8 \times 10^{-3}} \right)  \left(\frac{\dot M_{\rm peb}}{10^{-4} \rm  \Me \ yr } \right)^{-1}   \left(\frac{M_{\star}}{\Msun} \right)   \rm  \ yr }
     \hfill \hspace{1cm}   \mbox{[3D]},   \vspace{ 0.1 cm} 
     \end{cases}
   \end{multlined}
         \label{eq:t_PA}
\end{equation}
We can see that $t_{\rm PA}$ is proportional to $M_{\rm p}^{1/2}$  or $M_{\rm p}^{1/3}$ in the $2$D headwind or shear accretion regime,  while $\tau_{\rm PA}$ is independent of the planet mass in the $3$D regime.

\subsection{ Features}
 \label{sec:feature}
The first important feature is that pebble accretion is not a runaway process. From \eq{peb_accretion} we can see that $(dM/dt)/M \propto M^{0}$ in the $3$D regime, and $(dM/dt)/M \propto M^{-1/2}$ or $M^{-1/3}$ in the $2$D headwind or shear regime. This means that the mass ratios  among the growing bodies would approach the order of unity as the accretion proceeds. This is the feature of the orderly growth.  Nevertheless, even not in a runaway mode,  pebble accretion can still be very fast due to a large accretion cross section and a high flux of continuous feeding pebbles from the outer part of the disk  (see discussions in \se{comparison}).

The efficiency of pebble accretion is determined by the Stokes number of the pebbles.  When pebble accretion  is in the $3$D regime,  the efficiency increases with the Stokes number. This is because higher Stokes number pebbles sediment into a thinner vertical layer,  increasing the number density of pebbles for being  accreted.  When the accretion is in the $2$D regime, the efficiency decreases with $\tau_{\rm s}$. In this case,  higher Stokes number pebbles drift faster and have a lower probability to be accreted by the planet (see Fig.4 of \cite{Ormel2018}). The above dependences hold when pebbles are marginally coupled to the disk gas ($10^{-3} \lesssim \tau_{\rm s} \lesssim 1$). 
When $\tau_{\rm s}$ is much larger than the order of unity, gas drag is negligible and pebbles are more  aerodynamic like planetesimals. In that circumstance, the actual accretion rate drops substantially \citep{Ormel2010a}.  Conversely, when $\tau_{\rm s}\lesssim 10^{-3}$, the pebbles are tightly coupled to the gas flow. Accretion rate is also very low in this geometric regime \citep{Guillot2014}. Therefore, the preferred Stokes number for pebble accretion ranges from $10^{-3}$ to $1$, corresponding to $0.3$ mm to $30$ cm size particles at $5$ AU in the MMSN model.

Pebble accretion is suppressed when the disk is highly turbulent \citep{Morbidelli2015,Ormel2018,Rosenthal2018}. There are two reasons. First, pebbles are stirred up vertically due to the turbulent diffusion.    In a strongly turbulent case, a vertically extended distribution of pebbles leads to less amount of them being accreted, and the corresponding efficiency is lower. This is similar to the effect of the smaller pebbles in the $3$D regime discussed above.  Second, the pebble's random velocity also correlates with disk turbulence as $ \sqrt{\alpha_{\rm t} \taus} c_{\rm s}$ \citep{Ormel2007b}. The impact velocity between the pebble and planet is additionally contributed by this turbulent-induced motion.  The settling condition fails when the  turbulent velocity is very high (\se{criterion}), and therefore, the accretion is also significantly suppressed.  \cite{Ormel2018} obtained a pebble accretion efficiency formula by accounting for the stochastic turbulent velocity into the equation of motion for the pebble. Their results are in agreement with the pebble accretion measured from more realistic MHD simulations of \cite{Xu2017} as well as  the VSI hydrodynamic simulations of \cite{Picogna2018}.

We note that the  trajectories of pebbles  and accretion features could deviate from the above described paradigm that only  steady-state shear gas flow is considered  (\se{criterion}). There are two additional effects.  First, when  pebbles are sedimented and accreted onto the planet, the potential energy of pebbles is transferred into frictional heat, which raises the temperature of surrounding gas. The deep gas layer close to the planet is more dynamically convective caused by this accretion-driven heating, which may affect the pebble accretion.  Taking into account  both adiabatic and convective models of pebble accretion in hydrodynamic simulations, \cite{Popovas2018} found that even though an active mass mixing among different layers is indeed observed due to the vigorous gas motion, the net pebble accretion is not strongly affected, except for the smallest particles that are tightly coupled to the gas.

Second, when the pebbles fall into the region close to the planet with a temperature exceeding their evaporation temperature, these pebbles get vaporized, resulting in an enrichment of the planetary envelope rather than direct accretion onto the core \citep{Alibert2017b,Brouwers2018,Valletta2019}. This enrichment increases  the gas mean molecular weight of the envelope, resulting in a thinner, opaque envelope \citep{Venturini2016}.   Meanwhile, hydrodynamic simulations indicated that the envelopes of gas for low-mass protoplanets are not in a steady-state but rather get replenished by the surrounding disk gas \citep{Ormel2015,Fung2015,Cimerman2017}.  It is unclear how largely the envelope enrichment process would be affected by the above gas recycling.  The ablation of accreting pebbles with realistic radiative transfer plus gas replenishment models,  and how these affect the core mass growth and gas accretion are an active research topics for future investigations.
 
 \subsection{Applications}
 \label{sec:application}

{\bf Solar System} \\
The pebble accretion scenario has been used to explain the formation of the Solar System. Based on the fact that icy pebbles drift across the water-ice line and sublimate into small silicate pebbles, \cite{Morbidelli2015} inferred that the growth of  protoplanets is in a slow $3$D accretion interior to the ice line and a fast $2$D accretion exterior to the ice line. This results in low-mass progenitors of  terrestrial planets in the inner disk regions and massive cores of giant planets in the outer disk regions, in agreement with the  architecture of the Solar System.
\cite{Levison2015} found that before efficient pebble accretion commences, an early phase of velocity stirring of protoplanets  is required. Protoplanets then evolve into a bimodal mass distribution.  Due to dynamical friction, the massive protoplanets have low velocity dispersions. After this phase, only these massive protoplanets can undergo rapid pebble accretion and grow to giant planet cores. The above self-excitation process is essential to explain why only a few giant planet cores formed in the Solar System. If there were no such  pre-stirring phase, hundreds of Earth-mass objects would form instead \citep{Kretke2014}.  
\cite{Johansen2015} proposed the formation of asteroids and Kuiper belt objects by accreting chondrules, which are millimeter-sized spherules commonly found in primitive meteorites. 
Furthermore,  formation of the Galilean satellites has also been recently proposed to be aided by pebble accretion \citep{Shibaike2019,Ronnet2020}.

{\bf Exoplanetary Systems} \\
The pebble accretion scenario is also widely adopted to explain the formation of  super-Earths and gas gaints in exoplanetary systems. \cite{Lambrechts2014} constructed a dust growth and pebble drift model to investigate the formation of giant planet cores by pebble accretion. \cite{Bitsch2015} studied the influence of disk radial distance and evolutional phase on the growth and migration of a single protoplanet, and \cite{Johansen2019}  focused on  conditions under which growth can overcome migration. 
Numerous recent  studies have incorporated the pebble accretion model into an N-body code to investigate the impact of pebble accretion on final planetary architectures \citep{Matsumura2017,Lambrechts2019,Bitsch2019,Izidoro2019,Liu2019a,
Schoonenberg2019,Coleman2019,Wimarsson2020,Ogihara2020}.  For instance, \cite{Lambrechts2019} demonstrated that the final type of planetary system (terrestrial planets or super-Earths) is crucially determined by the pebble flux, or equivalently, the total mass of the  pebble reservoir in a protoplanetary disk.
 
{\bf Systems around Low-mass Stars} \\
 \begin{figure}
   \centering
   \includegraphics[scale=0.75, angle=0]{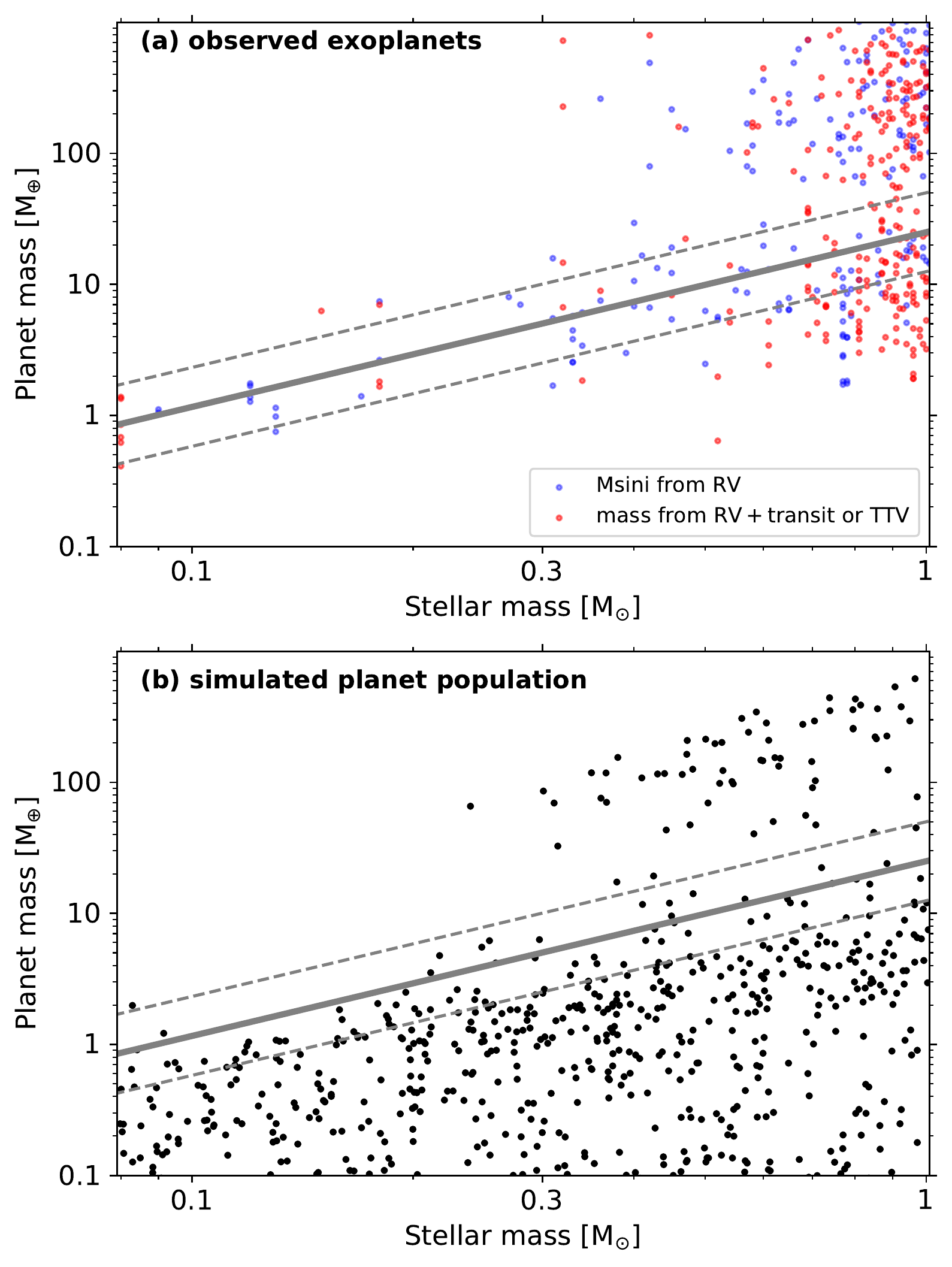}
   \caption{ Planet mass vs. stellar mass for the observed exoplanets (top) and simulated planet populations (bottom). 
Top panel: the blue dots are the planets only detected by RV surveys with a low mass limit, and the red dots are the planets with true masses either from  combined RV and transit surveys, or from transit timing variation measurements. The selected planets also have relatively precise stellar mass measurements ($\Delta \epsilon  {= }\Delta M_{\star}/M_{\star} {\leq}20\%$). The data are adopted from the NASA exoplanet archive, which is not corrected for  observational bias.  Bottom panel: Monte Carlo sampling plot of the resultant planets, where the initial protoplanetary seeds are assumed to form at the water-ice line  with masses of $0.01 \Me$. The solid line represents the characteristic mass of super-Earths, set by the pebble isolation mass from \eq{isomass}, whereas the dashed lines indicate a factor of two mass variation. The figure is reproduced based on \cite{Liu2019b}.
   }
   \label{fig:pps}
   \end{figure}
Studies of pebble accretion are not only limited to the systems around solar-mass stars, but also are generalized to stellar hosts of different masses.  \cite{Ormel2017} proposed a formation scenario for TRAPPIST-$1$ \citep{Gillon2016,Gillon2017} and other compact systems around very low-mass M-dwarf stars. In their scenario, protoplanets form by the streaming instability at the water-ice line. These protoplanets subsequently undergo inward migration and accrete pebbles to reach their final masses.  \cite{Ormel2017} suggested that  all TRAPPIST-$1$ planets of roughly Earth-mass could be an indication of the planet mass regulated by pebble isolation. The follow-up numerical simulations by \cite{Schoonenberg2019} verified that these forming planets have ${\approx}10\%$  water mass fractions, consistent with bulk density measurements and interior modeling of the TRAPPIST-$1$ planets \citep{Grimm2018,Unterborn2017,Dorn2018}.  

\cite{Liu2019b} investigated pebble-driven planet formation around stars of masses  from $0.08 \ \Msun$ to $1 \ \Msun$. 
\Fg{pps} illustrates the planet populations from  observations and the population synthesis model of \cite{Liu2019b}. It should be noted that the observed sample  is adopted from different surveys and uncorrected for any selectional bias. 
The figure nevertheless illustrates important features. 
First, a paucity of giant planets but not super-Earths  is found around stars with masses  below ${\simeq}0.1{-}0.2 \ \Msun$ (\Fg{pps}a).  Since larger planets are easier to detect than smaller planets for the same mass stellar hosts, the above planet desert is physical. Second, there seems to be a linear mass trend between the low-mass rocky-dominated planets (reflecting their core masses) and their stellar hosts for systems around stars less massive than $0.3 \ \Msun$. Again, the observational bias would not be the cause of this pattern, which leads to the opposite trend, due to the fact that more massive planets are easier to  detect around smaller stars. For systems around stars more massive than $ 0.3\Msun$,  we cannot directly infer the core masses, since the observed sample is outnumbered by  gas-rich giant planets. On the other hand, the above linear $M_{\rm c}{-} M_{\star}$  correlation exhibited in late-M dwarf systems is in good agreement with the inter-system uniformity reported from the super-Earth planets detected by the Kepler mission around  more massive early M-dwarfs and FGK stars \citep{Pascucci2018,Wu2019}. The validity of this positive correlation also needs to be unbiased and complementarily tested by current and future RV programs for a wider mass range of stellar hosts, such as High Accuracy Radial Velocity Planet Searcher, (HARPS, \citealt{Mayor2003, Mayor2011,Udry2019}), Echelle Spectrograph for Rocky Exoplanet and Stable Spectroscopic Observations (ESPRESSO, \citealt{Pepe2010}), Planet Finder Spectrograph, (PFS,\citealt{Crane2010,Feng2019}), and Calar Alto high-Resolution search for M dwarfs with Exoearths with Near-infrared and optical Echelle Spectrographs (CARMENES, \citealt{Quirrenbach2016}).

Importantly, \cite{Liu2019b} proposed that the characteristic mass (core mass) of super-Earths is set by the pebble isolation mass, which increases linearly with the mass of the stellar host. 
The pebble isolation mass can be written as \citep{Liu2019b} 
\begin{equation}
M_{\rm iso} = 25 \left( \frac{M_{\star}}{1 \ \Msun} \right)
\left( \frac{h_{\rm g} }{0.05} \right)^{3} \Me = 25 \left( \frac{M_{\star}}{1 \  \Msun} \right)^{4/3} \left( \frac{\dot M_{\rm g \odot}}{ 6 \times 10^{-8} \Msyr} \right)^{2/3}.
\label{eq:isomass}
\end{equation}
In the above equation, we neglect $\alphat$ and $\eta$ dependencies on $M_{\rm iso}$, $\dot M_{\rm g \odot}$ is the fiducial disk accretion rate  around the solar-mass star,  and $h_{\rm g}$ is derived from the viscously heated disk model by assuming that the disk accretion rate scales with the stellar mass squared.  \Fg{pps}(b) shows the resultant planets generated from the population synthesis model, and the solid line refers to the pebble isolation mass (\eq{isomass}). We clearly see that the super-Earths reaching $M_{\rm iso}$ increase with their stellar masses, approximately Earth-mass terrestrial planets around stars of $0.1 \ \Msun$ and $10{-}20 \ \Me$ around solar-mass stars. 
We also note that $M_{\rm iso}$ decreases  when the  disk evolves and $h_{\rm g}$ as well as  $\dot M_{\rm g}$ decline. For instance,   planets around solar-mass stars reach $M_{\rm iso}$ of $7 \Me$ when $\dot M_{\rm g \odot}{=}10^{-8} \Msyr$.
 Since $M_{\rm iso} $ is so low around late M-dwarfs,  massive gas giants are unlikely to form in such systems through the pebble accretion planet formation channel. \cite{Liu2020} further applied this approach for  even lower-mass central hosts and found the above linear mass scaling holds for planets around brown dwarfs.

\section{ Comparison between pebble accretion and planetesimal accretion}
\label{sec:comparison}
We compare the efficiency of pebble accretion and planetesimal accretion in \se{pebplt}. The stellar mass and radial distance dependence are discussed in \se{dependence}. We highlight the importance of incorporating these two accretion mechanisms for planet formation in \se{hybrid}. 
  
\subsection{ Why pebble accretion is more efficient than planetesimal accretion}
\label{sec:pebplt}
Pebble accretion has gained  attention since it is a more efficient growth process compared to  planetesimal accretion.  This comparison can be demonstrated  from the following two aspects: the accretion cross section and total mass of feeding materials.

For planetesimal accretion, the gravitational focusing factor reaches its maximum value when  $\delta v = v_{\rm H}$ (see \se{cross_section}).  Therefore,  the ratio between accretion radius and physical radius  is expressed as 
\begin{equation}
\begin{aligned}
b_{\rm PlA}/R  &=  \sqrt{  1 + \left( \frac{ {v_{\rm esc}}} {{R_{\rm H} \Omega_{\rm K}} } \right)^2} \simeq 2 \left(\frac{M_{\rm p}}{M_{\star}} \right)^{1/6} \left(\frac{a_{\rm }}{R_{\rm }} \right)^{1/2} \\
& \simeq 35  \left(\frac{a}{1 \rm \ AU} \right)^{1/2} \left( \frac{\rho_{\bullet}}{5 \ \rm gcm^{-3}} \right)^{1/2} \left( \frac{M_{\star}}{M_{\odot} }\right)^{-1/6}.
\label{eq:pl_enhance}
\end{aligned}
\end{equation}

In the shear-dominated  pebble accretion regime ($\Delta v \simeq b_{\rm peb} \Omega_{\rm K}$), from \eq{PA_radius} the accretion radius reads 
\begin{equation}
b_{\rm PA} \simeq \sqrt{ \frac{GM_{\rm p } t_{\rm stop}}{\Delta v }} \simeq  \tau_{\rm s}^{1/3} R_{\rm H}.
\label{eq:PA_shear}
\end{equation}
The pebble accretion radius can maximally approach the planet Hill radius when $\tau_{\rm s}$ reaches the  order of  unity. The enhanced factor is given by  
\begin{equation}
b_{\rm PA}/R  \simeq  \tau_{\rm s}^{1/3} \left(\frac{R_{\rm H}}{R} \right) \sim 230 \tau_{\rm s}^{1/3} \left(\frac{a}{1 \rm  \ AU}  \right) \left(\frac{\rho_{\bullet}}{5 \ \rm g \ cm^{-2}}  \right)^{1/3} \left( \frac{M_{\star}}{M_{\odot}}  \right)^{-1/3},
\label{eq:peb_enhance}
\end{equation}
 From \eqs{pl_enhance}{peb_enhance}, we can  see that generally the pebble accretion has a larger accretion cross section compared to the planetesimal accretion. This feature can be seen in \Fg{pltpeb}, which illustrates the trajectories of planetesimals (left) and pebbles (right) during their interactions with an Earth-mass planet.  Pebbles lose  angular momentum more efficiently and sediment towards to the planet  by a combination of gas drag and gravitational force. The above difference is more pronounced when the Stokes number of the pebbles is closer to the order of unity,  and the planet is further out and/or around less massive stellar hosts.
 
 \begin{figure}
   \centering
   \includegraphics[scale=0.75, angle=0]{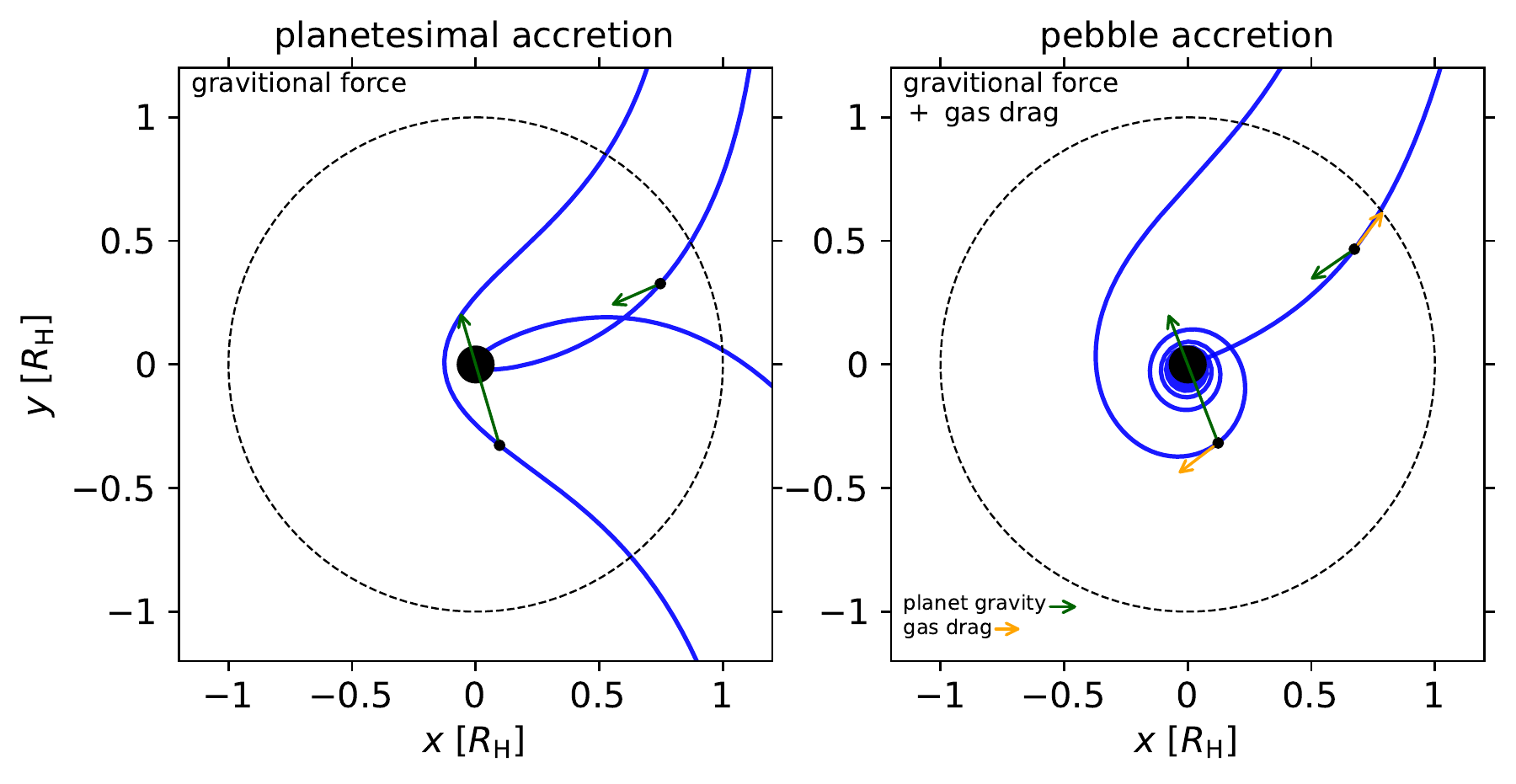}
   \caption{Illustration of planetesimal-planet (left) and pebble-planet (right) encounters,  viewed in a co-moving frame with a central planet.  The filled circle signifies an Earth-mass planet and the dashed circle indicates the planet Hill sphere. Two blue lines represent the trajectories of planetesimals or pebbles from the two different impact distances.
In the above illustration, planetesimals are deflected (left) while pebbles get accreted (right).  We note that  the central filled circle is not scaled with the physical size of the planet. 
  We set a zero-drag force for the planetesimal-planet interaction and pebbles of $\taus{=}1$ for the pebble-planet interaction, corresponding to ${\approx}10$ cm size  at $5$ AU in the MMSN model.
  The black dots represent the snapshots of planetesimals or pebbles, whereas the green and orange arrows correspond to the planet gravitational force and gas drag respectively. 
The direction and length of the arrow represent the direction and magnitude of the force respectively.  
   Aided by gas drag, pebbles can be more efficiently accreted by the planet compared to planetesimals.   
   }
   \label{fig:pltpeb}
   \end{figure}
 
 In addition, since planetesimals (${\sim} 100$ km) are weakly  affected by gas, the orbits of the planetesimals are relatively fixed. The planetary bodies only accrete their local planetesimals. The maximum  mass that the planet can reach (planetesimal isolation mass)  is written as $M_{\rm pl, iso} {= }2\pi a \Delta a \Sigma_{\rm pl}$ where the width of the feeding zone  $\Delta a {\sim} 10R_{\rm H}$.  For the MMSN,  the planet grows maximumly to Mars mass in the inner terrestrial  planet region and  a few Earth masses  beyond the water ice line (\eg, \citealt{Ida2004a}).    

On the contrary, pebble accretion is not limited to  the local pebble density.  The majority of the dust reside in the outer part of the disk.  Pebbles, which grow from dust grains, drift inwardly from the outer region of the disk.  Due to the mobility of pebbles, there is no concept of feeding zone for pebble accretion. The feeding materials  correspond to all pebbles that are able to bypass the orbit of the planet. The planets stop their core growth only when they reach the pebble isolation mass, which has a similar scaling as the gap opening mass (\eq{m_iso}). Therefore, in contrast to planetesimal accretion, the final core mass of the planet by pebble accretion mainly depends on the mass of the central star and the aspect ratio of the disk, but not the local density of solids.

\subsection{Stellar mass and radial distance dependence}
\label{sec:dependence}
Here we qualitatively discuss how these two accretion scenarios depend on the mass of the stellar host and the orbital distance.

First, we focus on the stellar mass dependence. Based on \eq{t_rg}, the growth timescale for planetesimal accretion is given by $t_{\rm PlA} \propto \Sigma_{\rm pl}^{-1} M_{\star}^{-1/2}$.  We have $t_{\rm PA} \propto \dot M_{\rm  peb}^{-1} M_{\star}^{1/2}$ in the $2$D headwind regime from \eq{t_PA}.  Since observations indicate   lower-mass disks  around less massive massive stars \citep{ Natta2006,Andrews2013,Pascucci2016}, the planetesimal surface density and pebble flux are expected to decrease with the mass of the stellar host. As a result, both pebble accretion and planetesimal accretion tend to be slower as $\Ms$ decreases. Furthermore,  assuming that the planetesimal surface density and pebble flux have the same stellar mass dependence, we find that the ratio of these two timescales  ($ \tau_{\rm PA}/ \tau_{\rm PlA}$) increases with $\Ms$. This in turn indicates that the decreasing rate of planetesimal accretion is faster than that of pebble accretion as $\Ms$ decreases. In other words,   pebble accretion is even more pronounced compared to planetesimal accretion for the planet growth around lower mass stars than higher mass stars. The above conclusion holds as well when adopting $t_{\rm PA}$  for the $2$D shear regime (\eq{t_PA}) or $t_{\rm PlA}$ corresponding to the  oligarchic growth  (\eq{t_og}).

The radial distance dependence is explained as follows.   Based on \eq{t_rg},  the planetesimal accretion timescale is written as $t_{\rm PlA} \propto \Sigma_{\rm pl}^{-1} r^{3/2}$. Since the disk surface density additionally decreases  with $r$,  $t_{\rm PlA}$ is expected to be a strongly increasing function of  $r$.  Planetesimals spend  a much longer time growing their masses at further out disk locations.  Moreover, as discussed in \se{plt_feature}, planet-planet encounters result in scatterings/ejections instead of collisions at large orbital distances. Therefore, the growth by planetesimal accretion is strongly suppressed or even quenched at large orbital distances. On the other hand,  for pebble accretion,  $ t_{\rm PA} \propto \eta$ in the $2$D shear regime. Although the growth by pebble accretion also turns out to be slower at further out disk locations, the radial distance dependence is weaker than planetesimal accretion.  Therefore, pebble accretion is more appealing  for the formation of distant massive planets.

To summarize, when comparing these two accretion scenarios, we find that pebble accretion becomes more attractive than  planetesimal accretion when the stellar host is less massive and/or the accretion occurs at a larger orbital distance.  

\subsection{ A hybrid accretion of pebbles and planetesimals }
  \label{sec:hybrid}
Despite the above distinctive differences, we want to raise the point that pebble accretion and planetesimal accretion are not two isolated, mutually exclusive growth channels. They are nevertheless likely to be connected and complementary for  planet growth. 
 
On the one hand, the above two mechanisms can operate concurrently and  jointly contribute to the mass increase. On the other hand, they also compete at certain levels since both pebbles and planetesimals are basic components of solids in protoplanetary disks \citep{Schoonenberg2019}.
 For instance, the streaming instability converts pebbles into planetesimals once the triggering condition is satisfied. After that, these forming planetesimals accrete surrounding planetesimals  as well as pebbles that continuously drift from the outer region of the disk. The following mass growth is  a combination of accreting pebbles and planetesimals \citep{Liu2019a,Schoonenberg2019}. We can consider two types of  extreme situations.  In one circumstance, the streaming instability is extensively triggered and the majority of the solids are in the form of planetesimals. The following growth is naturally led by planetesimal accretion. In the other circumstance, when the streaming instability is modestly triggered, the dominant solid masses are still in pebbles.  Therefore, pebble accretion is the central mechanism for the growth of the cores. As can be expected, a more general pattern is that  pebble and planetesimal accretions co-operate together to feed the planet growth.

In addition, the above two mechanisms may occur at different evolutionary stages \citep{Alibert2018,Venturini2020} and/or in different disk regions \citep{Ormel2017b}.  For example, when a planet reaches the pebble isolation mass, the inwardly drifting pebble flux is truncated. Pebble  accretion and planetesimal accretion occur for planets whose orbits are beyond the gap-opening planet. Interior to this massive planet, collisions between planetesimals and protoplanets are the only pathway for core mass growth.  This might provide an explanation for the mass dichotomy between the inner low-mass terrestrial planets and outer massive giant planets in the Solar System \citep{Ormel2017b}.  
All in all, the detailed exploration based on the concept of hybrid planetesimal and pebble accretion is an active research area and requires future studies.

\section{Summary and future outlook}
\label{sec:summary}

In this review, we have recapitulated the current states of exoplanet demographics and disk observations (\Se{introduction}). The planet formation theories have been overviewed chronologically, including dust coagulation and radial drift (\Se{dust}), planetesimal formation (\Se{streaming}), and subsequent planetesimal growth by planetesimal accretion (\Se{planetesimal}) and pebble accretion (\Se{PA}).
Importantly, we have discussed how different planet formation models fit into observations in each growth stage.

 Lastly, we  propose some open questions in this field, which are existing topics with disputed   interpretations and unsolved puzzles that require future studies.  These questions are summarized as follows:
\begin{itemize}
\item What is the characteristic size of solid particles in protoplanetary disks? Are these particles  mm-cm size as suggested by spectral index observations, or $100 \  \upmu$m as inferred from polarization measurements? On the one hand, the answer itself is valuable, since it needs to test the validity of different  model interpretations  and the built-in assumptions.  On the other hand,  the answer is also related to the subsequent planet formation processes such as  the streaming instability and pebble accretion, which essentially depend on the size of solid particles. \\

\item Current streaming instability and pebble acceretion studies are limited to the systems around single stars. Since the gas streamline in protoplantary disks around binaries significantly deviates from static, axisymmetric cases around single stellar hosts, it still remains to be seen how these gas-assisted formation mechanisms operate in this highly-perturbed binary environment.\\

\item How can we distinguish planetesimal accretion and pebble accretion mechanisms from an  observational perspective? Namely, what would be the key observational signatures that result from different formation channels (see \eg, \citealt{Brugger2020})? Can we really claim  which types of systems can be uniquely produced by pebble accretion, or vice versa? \\

\item Recent impact models found that the accreting pebbles as well as small planetesimals may get vaporized on their way to the planet interior due to the increased thermal ablation and friction (see \se{feature}).   The materials enrich the planetary atmospheres rather than directly impact the solid cores. The resultant core mass is substantially lower than the traditional critical mass that a planet requires for initiating rapid gas accretion \citep{Brouwers2020}.  How this mass deposition process affects the compositional and thermal structures of envelopes, and further influences the gas accretion are not well understood.  In addition, 
small bodies in protoplanetary disks feel head wind from the disk gas  and suffer the surface shear stress \citep{Paraskov2006,Schaffer2020}.
The influence of wind erosion during the above two accretion processes is also poorly investigated.  \\

\item How early can planets form? There seems to be evidence of large grains existing in the early Class I disk phase  \citep{Harsono2018}. Based on the isotope measurements of meteorites, the core of Jupiter with ${\sim}20 \Me$ is suggested to form early within $0.5{-}1$ Myr \citep{Kruijer2017}. Furthermore, if the rings and gaps in the HL Tau disk are induced by planets, it then indicates that planets of sub-Jupiter masses might already form at large orbital separations  within $1$ Myr in such a young system. Putting  all clues together, can we speculate that planet formation is more rapid than what we thought before?  \\

\item We have already learned much about of the planet-related properties from  exoplanet demographics: occurrence rate, mass, radius and corresponding dependencies on the mass and metallicity of the stellar host. The next step is to understand the planetary-system-related  properties, \ie, the architectures.  Can different populations of planets co-exist with each other under certain conditions,  or the formation of one type of planets inhibit the growth of the others? What would be the architecture of these systems? Some of these questions have already been pointed out by \cite{ZhuW2018b} and \cite{Masuda2020}.  In order to further address these questions, both follow-up observations and dedicated numerical modelings are needed.  \\

\item As raised by \cite{Murchikova2020}, is planet formation really an independent and generalized process  such that planets do not know each other and the environment where they reside? 
Or is planet formation  universally set by a few  key physical processes related to disk and host properties \citep{Liu2019b}, and therefore,  the imprinted planetary systems may contain some degrees of intra-system/inter-system similarity \citep{Millholland2017,WangSH2017,Weiss2018,Wu2019}?  \\ 
\end{itemize}

With a rapidly increasing number of characterized  exoplanets and protoplanetary disks,  we have greatly improved the statistics of planet formation, in the context of both initial conditions and final products.   The theoretical study of planet formation has  been advanced enormously in the last decade.  The validity of  modern planet formation scenarios needs to match various observational constraints.  Future work is also required to account for the consistency with findings from ongoing/upcoming space and ground-based facilities, such as TESS and ALMA.

\begin{acknowledgements}

We thank editor Wing-Huen Ip and the anonymous referee for useful comments and suggestions. We also thank Anders Johansen, Michiel Lambrechts, Joanna Dr\c{a}\.{z}kowska,  Chris Ormel, Rixin Li, Jiwei Xie, Shangfei Liu, Wei Zhu and Feng Long for proofreading the manuscript and providing  helpful  suggestions. B.L. greatly appreciates the valuable discussions with Gijs Mulders, Daniel Harsono, Joanna Dr\c{a}\.{z}kowska and Chris Ormel as a PPVII chapter proposal team.

In addition, B.L. wishes to express the deepest gratefulness to Adam Showman (1968-2020) for his inspiration and guidance in B.L.'s early academic career. Adam is an amazing scientist and a  tremendous mentor. His scientific contribution as well as  great personality will keep influencing and being remembered by the community.  Lastly, B.L. feels especially grateful to his girlfriend, Jing Yang, for her spiritual support during the writing period. In these tough COVID-19 pandemic times, many things that formerly seemed extraordinary have  now become ordinary. The work would not be done successfully without her dedicated encouragements.  

B.L. is supported by the European Research Council (ERC Consolidator Grant 724687-PLANETESYS), the Swedish Walter Gyllenberg Foundation, and start-up grant of Bairen program from Zhejiang University.  J.J.  is  supported by the B-type Strategic Priority Program of the Chinese Academy of Sciences (Grant No. XDB41000000), the National Natural Science Foundation of China (Grant Nos. 12033010 and 11773081), CAS Interdisciplinary Innovation Team and  Foundation of Minor Planets of the Purple Mountain Observatory.

\end{acknowledgements}

\begin{table*}
\centering
\begin{minipage}{4cm}
\caption{List of Notations}  \label{tb:long}  \end{minipage}

\fns
\begin{tabular}{ll}
\hline
{Symbol} & {Description} \\
\hline
     $v_{\rm g}  \ ( v_{\rm \phi, g}, v_{\rm r, g})$                &  gas velocity with azimuthal and radial components \\
      $v \ (v_{\rm \phi}, \ v_{\rm r})$ &  particle's velocity  with azimuthal and radial components  \\
      $v_{\rm peb} $ &  pebble velocity \\
      $v_{\rm K} $ & Keplerian velocity \\
      $v_{\rm th} $ &  gas mean thermal velocity \\
      $c_{\rm s} $ &  gas sound speed \\
       $ v_{\ast}$ & threshold velocity between settling and ballistic pebble accretion regimes  \\
       $ v_{\rm set}$ &  settling velocity of pebble onto planet    \\
      $v_{\rm esc} $ &  planet escape velocity   \\
      $\Delta v $ &  relative velocity between gas and particle \\
      $\delta v $ &  relative velocity between large and small planetesimals \\
      $\Omega_{\rm K} $ & Keplerian angular frequency \\
      $\eta $ &  gas headwind prefactor \\
      $P_{\rm g} $ & gas pressure \\
      $T_{\rm g} $ & gas temperature \\
       $ T_{\rm d}$ &  dust temperature    \\
      $H_{\rm g} $ &  gas disk scale height \\
      $H_{\rm peb} $ &  pebble disk scale height \\
      $H $ &  vertical height of small planetesimals \\
      $h_{\rm g} $ &  gas disk aspect ratio \\
      $h_{\rm peb} $ &  pebble disk aspect ratio \\
      $\Sigma_{\rm g} $ & gas surface density \\
      $\Sigma_{\rm d}$ &  dust surface density   \\
      $\Sigma_{\rm peb}$ & pebble surface density   \\
      $\rho_{\rm g} $ &  gas volume density \\
      $\rho_{\rm peb} $ &  pebble volume density \\
      $Z $ &  disk metallicity \\
      $Q_{\rm T} $ &  Toomre  parameter  \\
      $\gamma $ &  self-gravity parameter  \\
      $G $ &  gravitational constant \\
      $F_{\rm drag} $ &  hydrodynamic gas drag force   \\
      $a_{\rm drag} $ &  hydrodynamic gas drag acceleration  \\
      $C_{\rm D}  $ & coefficient of gas drag force  \\
      $R_{\rm e}  $ & particle's Reynolds number \\
      $\nu_{\rm mol}  $ &  kinematic molecular viscosity \\
      $\nu_{\rm t}  $ &  turbulent viscosity \\
      $\sigma_{\rm mol}  $ &   collisional cross section of  gas molecule \\
      $\lambda_{\rm mfp}  $ &  gas mean free path \\
      $m_{\rm mol}  $ &  mass of gas molecule \\
      $m_{\rm H}  $ &  mass of hydrogen atom  \\
      $\mu  $ &  gas mean molecular weight \\
      $\rho_{\bullet}$ &   internal density of solid body \\
      $r $ &   disk radial distance \\
      $R $ &   radius of solid body \\
      $R_{\rm H} $ &   Hill radius \\
       $a$ &  semimajor axis  \\
       $t_{\rm stop}$ &   stopping time   \\
       $\taus$ & Stokes number or  dimensionless stopping time  \\
       $\tau_{\nu} $ &  optical depth (at the corresponding  frequency $\nu$)  \\
      $I_{\nu} $ &  disk emission intensity   \\
       $ F_{\nu}$ &  observed flux   \\
       $\kappa_{\nu}$ &  disk opacity      \\
       $ B_{\nu}$ & Plank function   \\
       $ w_{\nu}$ & single-scattering albedo  \\
       $ c$ &  light speed     \\
       $\delta_{\rm d}$ & coefficient of gas diffusivity \\
        $\alpha_{\rm t}$ & efficiency factor of turbulent viscosity  \\
        $ \alpha$ &  spectral energy index      \\
        $ \beta$ &  dust opacity index      \\
        $ \beta_{\rm ISM}$ &  opacity index of ISM dust     \\
        $ \beta_{\rm disk}$ &  opacity index of protoplanetary dust     \\
        $ k_{\rm B}$ &  Boltzmann constant     \\
        $k_{\Sigma} $ &  gradient of disk surface density \\
        $k_{\rm T} $ &  gradient of disk temperature  \\
         $M_{\star} $ &  stellar mass \\
        $ M_{\rm pl}$ &  characteristic mass of planetesimal     \\
        $ M_{\rm p}$ &  planet mass  \\
        $ M$, $m$ &  mass of large and small planetesimals  \\
        $ R_{\rm M}$, $R_{\rm m}$ &  radius of large and small planetesimals  \\
        $ \sigma$ &  collisional cross section of large planetesimal   \\

\hline
\end{tabular}
\end{table*}

\setcounter{table}{0}
\begin{table*}
\centering
\begin{minipage}{3cm}
\caption{\it Continued.} 

\end{minipage}

\fns
\begin{tabular}{ll}
\hline
{Symbol} & {Description} \\
\hline
$ n_{\rm m}$, $ n_{\rm M}$ &  number density of small and large planetesimals  \\
        $ \Sigma_{\rm m}$,  $ \Sigma_{\rm M}$ &  surface density of small and large planetesimals  \\
        $ t_{\rm set}$ &  settling timescale of pebble onto planet    \\
       $ t_{\rm enc}$ &  pebble-planet encounter timescale \\
         $t_{\rm PA}$  & pebble accretion timescale  \\
         $t_{\rm PlA}$  & planetesimal accretion timescale  \\
        $ t_{\rm grow}$ &  mass growth timescale   \\
        $ t_{\rm vs}$ &  viscous stirring timescale   \\
        $ t_{\rm e, gas}$ &  gas eccentricity damping  timescale   \\
      $ t_{\rm og}$ &  oligarchic growth  timescale   \\
      $ t_{\rm rg}$ &   runaway growth  timescale   \\
      $ C_{\rm rg}$ &   numerical prefactor of runaway growth  \\
      $\Phi $ &  ratio between planetary escape velocity and Keplerian velocity   \\
      $ \ln \Lambda $ &   Coulomb factor   \\
       $ b$ &  impact parameter during pebble-planet encounter  \\
       $ b_{\rm set}$ & pebble accretion radius in the settling regime  \\
       $ b_{\rm PA}$ & pebble accretion radius \\
       $ b_{\rm PlA}$ & planetesimal accretion radius \\
        $ f_{\rm set}$ & fitting function between settling and ballistic regimes   \\
        $ f_{\rm gf}$ & gravitational focusing factor  \\
        $M_{\rm onset}$ &  planet mass for  onset of  efficient pebble accretion  \\
        $M_{\rm iso}$ &  planet mass for  termination of pebble accretion \\
        $M_{\rm hw/sh}$ &  planet mass between headwind and shear-dominated pebble accretion regimes \\
        $\dot M_{\rm g}$ & gas disk accretion rate  \\
        $\dot M_{\rm g \odot}$ & gas disk accretion rate around solar-mass star \\
        $\dot M_{\rm peb}$ & disk pebble flux  \\
        $\dot M_{\rm PA}$ & pebble accretion rate \\
        $\varepsilon_{\rm PA}$  & pebble accretion efficiency  \\
\hline
\end{tabular}
\end{table*}

%\bibliographystyle{raa.bst}
%\bibliography{reference.bib}

%\bibliographystyle{/home/lixh/raa/bibstyle/raa}
%\bibliography{bibtex}

\end{document}